\documentclass[nohyperref]{article}

\usepackage{microtype}
\usepackage{graphicx}
\usepackage{subfigure}
\usepackage{booktabs} 

\usepackage{hyperref}

\usepackage[accepted]{icml2022}

\usepackage{amsmath}
\usepackage{amssymb}
\usepackage{mathtools}
\usepackage{amsthm}

\usepackage[capitalize,noabbrev]{cleveref}

\theoremstyle{plain}
\newtheorem{theorem}{Theorem}[section]

\newtheorem{corollary}[theorem]{Corollary}
\theoremstyle{definition}

\theoremstyle{remark}

\usepackage[textsize=tiny]{todonotes}

\usepackage{amsfonts}
\usepackage{amsmath}
\usepackage{mathrsfs}  
\usepackage{amssymb}
\usepackage{graphicx}
\usepackage{booktabs}
\usepackage{enumitem}

\DeclareUnicodeCharacter{0229}{\c{a}}

\usepackage{mathtools}

\DeclareMathOperator{\argmin}{argmin}

\usepackage{xparse}

\usepackage{amsthm}

\numberwithin{observation}{section}

\icmltitlerunning{Correlation Clustering via Strong Triadic Closure Labeling}

\begin{document}

\twocolumn[
\icmltitle{Correlation Clustering via Strong Triadic Closure Labeling: Fast Approximation Algorithms and Practical Lower Bounds}



\icmlsetsymbol{equal}{*}

\begin{icmlauthorlist}
\icmlauthor{Nate Veldt}{yyy}
\end{icmlauthorlist}

\icmlaffiliation{yyy}{Department of Computer Science and Engineering, Texas A\&M University, College Station, Texas, USA}

\icmlcorrespondingauthor{Nate Veldt}{nveldt@tamu.edu}

\icmlkeywords{Correlation clustering, strong triadic closure, approximation algorithms}

\vskip 0.3in
]



\printAffiliationsAndNotice{}  

\begin{abstract}
Correlation clustering is a widely studied framework for clustering based on pairwise similarity and dissimilarity scores, but its best approximation algorithms rely on impractical linear programming relaxations. We present faster approximation algorithms that avoid these relaxations, for two well-studied special cases: cluster editing and cluster deletion. We accomplish this by drawing new connections to edge labeling problems related to the principle of strong triadic closure. This leads to faster and more practical linear programming algorithms, as well as extremely scalable combinatorial techniques, including the first combinatorial approximation algorithm for cluster deletion. In practice, our algorithms produce approximate solutions that nearly match the best algorithms in quality, while scaling to problems that are orders of magnitude larger.
\end{abstract}

\section{Introduction}
Correlation clustering is a framework for unsupervised learning that clusters items in a dataset based on pairwise similarity and dissimilarity scores, rather than based on explicit representations for data objects.
The simplest version of the problem can be cast as a partitioning objective on a complete signed graph, where the goal is to form clusters of nodes in a way that avoids placing negative edges inside clusters or positive edges between clusters. 
Bansal, Blum, and Chawla~\cite{BansalBlumChawla2004} introduced the problem, proved its NP-hardness, and presented the first approximation algorithms.  Now, nearly two decades later, developing improved approximation algorithms and hardness results for different variants of the problem continues to be a very active area of research~\cite{jafarov2020ccasymmetric,jafarov2021local,veldt2020parameterized,bun2021differentially,cohen-addad2021correlation,bonchi2022correlation}.
In addition to extensive theoretical research, the problem has been applied in a wide variety of settings, including to image segmentation~\cite{kim2011highcc,yarkony2012fast}, community detection~\cite{wang2013scalable,Veldt:2018:CCF:3178876.3186110,chen2012clustering}, cross-lingual link detection~\cite{van2007correlation}, cancer mutation analysis~\cite{hou2016cancercc}, and detecting co-regulated genes based on expression profiles~\cite{bhattacharya2010average,bhattacharya2008divisive,Ben-DorShamirYakhini1999}. However, despite significant previous research on both theoretical and applied aspects of correlation clustering, there still exists a wide gap between the best approximation algorithms and the most practical tools for this task. Many fast heuristic methods come with no approximation guarantees~\cite{shi2021scalable,Beier_2015_CVPR,bhattacharya2008divisive,levinkov2017comparative}, while rigorous approximation algorithms are often either impractically expensive, or only apply to the simplest version of the problem.

\emph{Linear programming algorithms.} The best approximation factors for correlation clustering, both for the standard unweighted objective~\cite{ChawlaMakarychevSchrammEtAl2015} as well as more general weighted variants~\cite{vanzuylen2009deterministic,AilonCharikarNewman2008,puleo2015correlation,puleo2018correlation,Veldt:2018:CCF:3178876.3186110,gleich2018ccgen,li2019motif,jafarov2020ccasymmetric}, are obtained by solving an expensive linear programming (LP) relaxation in order to obtain lower bounds for the objective. 
These lower bounds can useful for more than just designing approximation algorithms. In practice, linear programming lower bounds are often much closer to optimality than worst case theoretical results predict. If a good LP lower bound can be computed, the output of fast heuristic methods can be compared against the lower bound to obtain a posteriori approximation guarantees that are often very good in practice~\cite{swoboda2017message,lange2018partial,yarkony2012fast,Veldt:2018:CCF:3178876.3186110}. However, despite some recent work on specialized solvers~\cite{veldt2019metric,ruggles2020parallel,sonthalia2020project}, these lower bounds can only be computed for medium-sized instances at best (e.g., graphs with a few thousands nodes), and even then this can take a long time. There is therefore a need for faster approximation algorithms, as well as an even more basic need to efficiently compute good lower bounds for the NP-hard objective in practice. 

\emph{Pivot algorithms.} \textsc{Pivot} is a combinatorial algorithm that iteratively selects a random unclustered node and places it with all of its unclustered positive neighbors. This provides a randomized 3-approximation for complete unweighted correlation clustering~\cite{AilonCharikarNewman2008}. 
This approach can be made very fast~\cite{xinghao2015parallel,chierichetti2014correlation,cohen-addad2021correlation}, but is not without its limitations. First, the approach is designed for the complete unweighted case and does not extend as easily as LP-based techniques to other variants of correlation clustering. As one example, \emph{cluster deletion} is a simple variant that strictly prohibits clustering two nodes together if they share a negative edge. Although constant-factor linear programming algorithms have been designed for cluster deletion~\cite{puleo2015correlation,Veldt:2018:CCF:3178876.3186110,CharikarGuruswamiWirth2005}, the standard \textsc{Pivot} technique does not even produce a feasible solution for this problem.
Another limitation of \textsc{Pivot} is that it only provides an expected approximation guarantee, and although derandomization techniques have been developed, these again rely on solving the expensive canonical LP relaxation~\cite{vanzuylen2009deterministic}. Finally, \textsc{Pivot} does not produce explicit lower bounds for the NP-hard objective, so it provides no way to compute improved a posteriori approximation guarantees in practice.

\textbf{The present work: practical lower bounds and faster algorithms via connections to strong triadic closure.}
We provide significant steps in bridging the theory-practice gap in correlation clustering by designing practical techniques for computing lower bounds, and faster corresponding approximation algorithms. 
We achieve our results by highlighting and exploiting a relationship between correlation clustering and edge-labeling problems related to the principle of \emph{strong triadic closure}~\cite{sintos2014using,easley2010networks}. Strong triadic closure (STC) posits that two people will often share at least a weak connection if they share strong connections to a mutual friend. An STC-labeling of a graph is a way to label edges as weak or strong in order to satisfy this principle. Similarities between clustering and STC-labeling problems have been noted in previous work~\cite{konstantinidis2018strong,gruttemeier2020relation}. Our results significantly expand on these previously observed relationships, and provide new strategies for rounding lower bounds for STC-{labeling} problems into approximate solutions for clustering problems.


One of the central contributions of our paper is to develop \emph{combinatorial} approximation algorithms for complete unweighted correlation clustering (also called cluster editing) and cluster deletion, which can also be made {deterministic} without LP relaxations. Our strategy works in three basic steps: \emph{matching}, \emph{flipping}, and \emph{pivoting}. We first obtain lower bounds by computing maximal matchings in either an auxiliary graph (for cluster deletion) or an auxiliary 3-uniform hypergraph (for cluster editing). We use the results of our matching to determine edges to be flipped (i.e., added or deleted) to create a new graph. Finally we prove that applying a pivoting procedure on the resulting graph yields approximation guarantees for the original problem. 
In summary, we provide the following algorithms:
\vspace{-.75em}
\begin{itemize}[leftmargin=5pt, itemsep = -0.25em]
	\item We apply our \emph{match-flip-pivot} approach to design the first combinatorial approximation algorithm for cluster deletion, which provides a 4-approximation guarantee.
	\item We use \emph{match-flip-pivot} to design a combinatorial 6-approximation algorithm for complete unweighted correlation clustering. This can be made deterministic using purely combinatorial techniques, making it the best approximation guarantee achieved by any deterministic algorithm that does not depend on linear programming.
	\item We present additional approximation algorithms for both problems, each with a factor-4 approximation guarantee, by solving and rounding LPs with significantly fewer constraints than the canonical LP relaxations.
\end{itemize}
\vspace{-.75em}

In proving these results, we show more generally that any $\alpha$-approximation algorithm for the \emph{minimum weakness strong triadic closure} problem~\cite{sintos2014using} can be used to design a $(2\alpha)$-approximation for cluster deletion. We show an analogous result for unweighted complete correlation clustering. 
Our results imply that an optimal solution to either of these clustering problems is always within a factor of 2 of the optimal solution to a corresponding edge labeling problem. This is especially significant for cluster deletion and minimum weakness strong triadic closure, as there are known examples where these objectives differ by up to a factor of $8/7$~\cite{gruttemeier2020relation}. Thus, our upper bound on the difference between these two objectives is not far off of a known lower bound on their difference. Finally, we show that our algorithms are far more scalable in practice than algorithms that solve the canonical LP, producing approximate solutions that are within a small factor of optimality (factors $\approx 2$) on graphs with millions of nodes, within a matter of seconds. Our fast algorithms for finding lower bounds can also be used to provide good a posteriori approximation guarantees for heuristics that previously came with no guarantees.


\section{Preliminaries and Related Work}
We begin with definitions, terminology, and notation. 
Although correlation clustering is often presented as a clustering objective on a signed graph, the variants we primarily consider can be cast as objectives on an unsigned and unweighted graph $G = (V,E)$. All objectives we consider in this paper are defined on undirected graphs.

\textbf{Open wedges.} 
The problems we consider rely on the notion of an \emph{open wedge}. Given an unsigned graph $G = (V,E)$, a triplet of nodes $(i,j,k)$ is an \emph{open wedge centered at $k$} if $(i,k) \in E$ and $(j,k) \in E$ but $(i,j) \notin E$. Let $\mathcal{W}$ denote the set of triplets that define open wedges, and $\mathcal{W}_k \subseteq \mathcal{W}$ denote the set of open wedges centered at $k$. 


\subsection{Correlation Clustering Objectives}
The most general weighted version of correlation clustering is defined by a node set $V$ and two sets of nonnegative weights $W^+$ and $W^-$ defined on pairs of nodes. Formally, each pair $(i,j) \in {V \choose 2}$ is associated with an edge weight $w_{ij}^+ \in W^+$ and an edge weight $w_{ij}^- \in W^-$. In some applications these correspond to edge weights for positive and negative edges in a signed graph, but this does not always have to be the case.
Broadly speaking, the nonnegative weights $(w_{ij}^+, w_{ij}^-)$ indicate the level of attraction and repulsion, respectively, between nodes $i$ and $j$ in a clustering problem. The goal of the general weighted correlation clustering objective is to partition the nodes in a way that correlates as much as possible with these weights. This is accomplished by solving the following integer linear program (ILP), know as the \emph{MinDisagree} objective:
\begin{equation}
	\label{genccilp}
	\begin{array}{ll}
		\min & \displaystyle{\sum_{(i,j) \in {V \choose 2}}} w_{ij}^+ x_{ij} + w_{ij}^- (1 - x_{ij}) \\
		\text{such that} & x_{jk} + x_{ik} \geq x_{ij} \text{ for triplets $i,j,k$} \\
		&  x_{ij} \in \{0,1\} \text{ for $(i,j) \in {V \choose 2}$}. \\
	\end{array}		
\end{equation}
Here, $x_{ij} = 0$ if nodes $i$ and $j$ are clustered together and $x_{ij} = 1$ if they are separated. In other words, a penalty of $w_{ij}^+$ is applied for separating $i$ and $j$, and a penalty of $w_{ij}^-$ is applied if they are placed together. Note that
there is a triangle inequality constraint $x_{ij} \leq x_{jk} + x_{ik}$ for each ordering of three distinct vertices $\{i,j,k\}$. An alternative objective for correlation clustering is to maximize the weight of agreements, which is the same at optimality but is different from the perspective of approximations. Throughout this paper, we focus on the \emph{MinDisagree} objective~\eqref{genccilp}.

Correlation clustering is NP-hard even for the simple unweighted case, but if the binary constraint $x_{ij} \in \{0,1\}$ is relaxed to linear constraints $0 \leq x_{ij} \leq 1$, the result is the canonical linear programming relaxation of the problem, which can be solved in polynomial time.
Many approximation algorithms rely on solving and rounding this LP. An $O(\log n)$ approximation can be obtained for the general case using LP rounding~\cite{CharikarGuruswamiWirth2005,DemaineEmanuelFiatEtAl2006}, and improved results exist for special weighted variants~\cite{Veldt:2018:CCF:3178876.3186110,puleo2015correlation,jafarov2020ccasymmetric,AilonCharikarNewman2008,veldt2020parameterized}.

\textbf{Cluster editing.} 
Minimizing disagreements in a complete unweighted signed graph~\cite{BansalBlumChawla2004} is the widely-studied special case of objective~\eqref{genccilp} where $(w_{ij}^+, w_{ij}^-) \in \{(0,1), (1,0)\}$ for each node pair $(i,j)$. This is equivalent to a problem called \emph{cluster editing}: given a graph $G = (V,E)$, find the minimum number of edges to add or remove in order to convert $G$ into a disjoint union of cliques~\cite{ShamirSharanTsur2004,Ben-DorShamirYakhini1999}. Alternatively, this means clustering $G$ in a way that minimizes the number of \emph{mistakes} or \emph{disagreements}: edges crossing between clusters or non-adjacent node pairs inside clusters.

\textbf{Cluster deletion.} 
Cluster deletion~\cite{ShamirSharanTsur2004} seeks to convert a graph $G = (V,E)$ into a disjoint union of cliques by deleting the smallest number of edges. This can be viewed as an instance of general weighted correlation clustering~\eqref{genccilp} where $(w_{ij}^+, w_{ij}^-) = (1,0)$ if $i$ and $j$ are adjacent in $G$, and $(w_{ij}^+,w_{ij}^-) = (0, \infty)$ if these nodes are not adjacent. In this special case, the problem permits an integer programming formulation with fewer constraints, since $x_{ij} = 1$ when $(i,j) \notin E$:
\begin{equation}\label{cdilp}
	\begin{array}{lll}
		\min & \sum_{(i,j) \in E} x_{ij}  &\\
		\text{s.t.} & x_{ij} \in \{0,1\} & \text{for all $(i,j) \in E$} \\
		
&		x_{jk} + x_{ik} \geq 1 & \text{if $(i,j,k) \in \mathcal{W}_k$} \\ 
		&	\begin{rcases}
			x_{jk} + x_{ik} \geq x_{ij} \\ 
			x_{jk} + x_{ij} \geq x_{ik} \\
			x_{ik} + x_{ij} \geq x_{jk}  \\
		\end{rcases}	& \text{ if $i,j,k$ is a triangle.}
	\end{array}		
\end{equation}
The best approximation factor for cluster deletion is 2~\cite{Veldt:2018:CCF:3178876.3186110}, which
relies on rounding the linear programming relaxation of objective~\eqref{cdilp}.

\subsection{Strong Triadic Closure Labeling Objectives}
Strong triadic closure~\cite{easley2010networks,granovetter1973strength} posits that two people are likely to share at least a weak connection if they both share strong connections to a mutual friend. 
This is used as a guiding principle for social network analysis, and is the foundation for certain edge labeling problems~\cite{sintos2014using,gruttemeier2020relation,konstantinidis2018strong}.


\textbf{Min-weakness strong triadic closure.}
If every edge in a graph $G = (V,E)$ is labeled as either \emph{weak} or as \emph{strong}, we say this is a \emph{strong triadic closure labeling} if at least one of the edges in each open wedge is \emph{weak}. The rationale is that if both edges in a wedge are {strong} ties, we would expect the wedge to be closed because of strong triadic closure. The  \emph{minimum weakness strong triadic closure} (\textsc{minSTC}) problem~\cite{sintos2014using} seeks a strong triadic closure labeling with the minimum number of weak edges. We can cast this as an integer program where a binary variable $z_{uv}$ equals 1 if $(u,v)\in E$ is labeled as a weak connection:
\begin{equation}
	\label{minstc}
	\begin{array}{lll}
		\min & \sum_{(i,j) \in E} z_{ij} &\\
		\text{s.t. } & z_{jk} + z_{ik} \geq 1 &\text{ if $(i,j,k) \in \mathcal{W}_k$}\\
		&  z_{ij} \in \{0,1\} &\text{ for all $(i,j) \in E$}.
	\end{array}		
\end{equation}
%
\textbf{Strong triadic closure with edge additions.}
Minimum-weakness strong triadic closure with edge additions (\textsc{minSTC+}) allows one to satisfy strong triadic closure by also \emph{adding} weak edges between non-adjacent nodes in the graph $G = (V,E)$. This is equivalent to viewing certain \emph{non-edges} as weak connections that were not observed. The integer program formulation for this problem is:
\begin{equation}
	\label{mine}
	\begin{array}{lll}
		\min & \displaystyle{\sum_{(i,j) \in {V \choose 2}}} z_{ij} &\\
		\text{s.t. } & z_{jk} + z_{ik} + z_{ij} \geq 1 &\text{ if $(i,j,k) \in \mathcal{W}$}\\
		&  z_{ij} \in \{0,1\} &\text{ for $(i,j) \in {V \choose 2}$}.
	\end{array}		
\end{equation}
If $(i,j) \in E$ and $z_{ij} = 1$, this again corresponds to labeling the edge as weak. If $(i,j) \notin E$ and $z_{ij} = 1$, this means we \emph{add} a new edge between $i$ and $j$. A feasible solution to \textsc{minSTC+} is therefore a set of new edges $E'$ and a subset of edges $E_W \subseteq E$ that we will label as \emph{weak}. The goal is to minimize $|E'| + |E_W|$. We assume all new edges $E'$ are weak, to ensure we do not introduce new open wedges that violate strong triadic closure. We refer to a feasible solution $(E', E_L)$ for \textsc{minSTC+} as an \textsc{STC+} labeling. 

\subsection{Correlation Clustering and STC Connections}
The cluster deletion~\eqref{cdilp} integer program can be obtained by taking the integer program for \textsc{minSTC}~\eqref{minstc} and adding the constraints $z_{ij} + z_{ik} \geq z_{jk}$ for all permutations of nodes $i,j,k$ when $(i,j,k)$ is a triangle in $G$. Thus, a feasible solution for cluster deletion will produce a feasible solution for \textsc{minSTC} by labeling all the deleted edges as weak. Similarly, the constraints in the integer program for \textsc{minSTC+}~\eqref{mine} can be seen as a subset of the triangle inequality constraints in the integer program for cluster editing, once a change of variables is applied. The fact that \textsc{minSTC} lower bounds cluster deletion, and \textsc{minSTC+} lower bounds cluster editing, has already been noted in previous work~\cite{gruttemeier2020relation,gruttemeierstrong,konstantinidis2018strong}. The optimal solutions to cluster deletion and \textsc{minSTC} are known to coincide for co-graphs~\cite{konstantinidis2018strong}, though there exist concrete examples to confirm that in general they are not the same problem~\cite{gruttemeier2020relation}. There also exist approximation algorithms for correlation clustering that rely on lower bounds from open wedge packings (also called \emph{bad triangle packings})~\cite{AilonCharikarNewman2008,BansalBlumChawla2004}. These can also be viewed implicitly as lower bounds for related STC-labeling problems, though this connection is not explicitly addressed in these works. Our paper expands on the known relationship between correlation clustering and STC-labeling problems, and provides new ways to round lower bounds for labeling problems into approximate solutions for clustering problems.

\section{Faster Linear Programming Algorithms}
We first show how to round LP relaxations for STC-labeling problems to develop 4-approximation algorithms for cluster deletion and cluster editing. While improved guarantees can be obtained by rounding tighter relaxations~\cite{ChawlaMakarychevSchrammEtAl2015,Veldt:2018:CCF:3178876.3186110}, our approach provides a useful tradeoff in runtime and approximation guarantee. In our experiments, we find that these less expensive LP relaxations are faster, while typically performing just as well as algorithms based on canonical relaxations. To prove our results, we first review a general pivoting strategy for correlation clustering, which we build on in our work.

\subsection{Algorithmic Pivoting Primitive}
\textsc{Pivot} provides an expected 3-approximation for cluster editing~\cite{AilonCharikarNewman2008}, but applying it directly to a weighted graph typically yields poor results. For cluster deletion, it does not even produce a feasible solution. However, pivoting can be successfully used as a {step} in more sophisticated algorithms for correlation clustering. 
We extract a general algorithmic strategy from the work of van Zuylen and Williamson~\cite{vanzuylen2009deterministic}. This method takes in a weighted instance of correlation clustering $(V, W^+, W^-)$, a set of ``budgets'' $\{b_{ij} \colon { (i,j) \in {V \choose 2}} \}$, and a derived graph $\hat{G} = (V,\hat{E})$. Theorem~\ref{thm:3pt1} provides conditions for bounding the output solution from running \textsc{Pivot} on $\hat{G}$.
\begin{theorem}
	\label{thm:3pt1} 
	(Thm 3.1,~\citet{vanzuylen2009deterministic}.)
	Let $(V, W^+, W^-)$ define a weighted instance of correlation clustering~\eqref{genccilp}, and $b_{ij}$ define the \emph{budget} for node pair $(i,j) \in {V \choose 2}$. Assume that for some $\alpha > 0$, there is a graph $\hat{G} = (V,\hat{E})$ satisfying the following two properties:
	\begin{enumerate} 
		\item For all~$(i,j) \in \hat{E}$, we
		have~$w_{ij}^- \leq \alpha b_{ij}$, and for all $(i,j) \notin \hat{E}$, we have $w_{ij}^+ \leq \alpha b_{ij}$.
		\item If $(i,j,k)$ is an open wedge centered at j in $\hat{G}$, we have $w_{ij}^+ + w_{jk}^+ + w_{ik}^- \leq \alpha \left(   b_{ij} + b_{jk} + b_{ik} \right)$.
	\end{enumerate}
	Then 
	applying \textsc{Pivot} to $\hat{G} = (V,\hat{E})$ with uniform random pivots will return a clustering with expected weight of disagreements bounded by $\alpha \sum_{i < j} b_{ij}$. There also exists a deterministic pivoting strategy that returns a clustering with disagreements bounded above by  $\alpha \sum_{i < j} b_{ij}$. 
\end{theorem}
Appendix~\ref{app:det} provides details for the deterministic strategy. By setting budgets $b_{ij}$ to be the contribution of a node pair $(i,j)$ to the LP relaxation of~\eqref{genccilp}, i.e., $b_{ij} = w_{ij}^+ x_{ij} + w_{ij}^- (1-x_{ij})$, one can obtain a derandomized 3-approximate \textsc{Pivot} method for complete unweighted correlation clustering~\cite{vanzuylen2009deterministic}. However, the bottleneck is solving the LP relaxation.  

\subsection{Faster LP Algorithms for Clustering}
\label{labelinglps}
We first present a new approximation algorithm for cluster editing by rounding the LP relaxation of \textsc{minSTC+}~\eqref{mine}, obtained by replacing $z_{ij} \in \{0,1\}$ with linear constraints $0 \leq z_{ij} \leq 1$. We apply a convenient change of variables: set $x_{ij} = z_{ij}$ if $(i,j) \in E$, and $x_{ij} = 1-z_{ij}$ otherwise.
This leads to a linear program that amounts to the canonical cluster editing LP relaxation but with fewer constraints:
\begin{equation}\label{owlx}
\begin{array}{ll}
\min & {\sum_{(i,j) \in E} x_{ij} + \sum_{(i,j) \notin E} (1-x_{ij})} \\
\text{s.t. } &  x_{ij} \leq x_{jk} + x_{ik} \text{ if $(i,j,k) \in \mathcal{W}_k$}\\
& 0 \leq x_{ij} \leq 1 \text{ for all $(i,j) \in {V \choose 2}$}.
\end{array}		
\end{equation}
The number of constraints in this LP is $O(|V|^2 + |\mathcal{W}|)$. Assuming a connected graph, this is bounded above by $O(|E||V|)$ and for many common graph classes will be smaller than the $O(|V|^3)$ constraints required for the canonical relaxation.
\begin{algorithm}[tb]
	\caption{Rounding the \textsc{minSTC+} LP relaxation.}
	\begin{algorithmic}[5]
		\STATE{\bfseries Input:} Graph $G = (V,E)$
		\STATE {\bfseries Output:} Clustering of $G$.
		\STATE Solve LP~\eqref{owlx}
		\STATE Set $\hat{E} \gets \{(i,j) \in {V \choose 2} : x_{ij} < 1/2 \}$
		\STATE Return $\textsc{Pivot}(\hat{G} = (V,\hat{E}))$
	\end{algorithmic}
	\label{alg:4owl}
\end{algorithm}
Algorithm~\ref{alg:4owl} is a 4-approximation for cluster editing. To prove its approximation guarantee, we show how to round the output of LP~\eqref{owlx} to produce a graph satisfying the conditions of Theorem~\ref{thm:3pt1} with $\alpha = 4$. 
\begin{theorem}
	\label{thm:4ce}
	Algorithm~\ref{alg:4owl} is a randomized 4-approximation algorithm for cluster editing.
\end{theorem}
Algorithm~\ref{alg:4cd} rounds the~\textsc{minSTC} LP relaxation, obtained by replacing $x_{ij} \in \{0,1\}$ with $0 \leq x_{ij} \leq 1$ in~\eqref{minstc}.
\begin{algorithm}[tb]
	\caption{Rounding the \textsc{minSTC} LP relaxation.}
	\begin{algorithmic}[5]
		\STATE{\bfseries Input:} Graph $G = (V,E)$
		\STATE {\bfseries Output:} Feasible cluster deletion clustering of $G$.
		\STATE Solve LP relaxation of~\eqref{minstc} 
		\STATE Set $\hat{E} \gets \{(i,j) \in E: z_{ij} < 1/2 \}$
		\STATE Return $\textsc{Pivot}(\hat{G} = (V,\hat{E}))$
	\end{algorithmic}
	\label{alg:4cd}
\end{algorithm}
\begin{theorem}
	\label{thm:4cd}
	Algorithm~\ref{alg:4cd} is a randomized 4-approximation algorithm for cluster deletion.
\end{theorem}
The appendix provides full proofs for Theorems~\ref{thm:4ce} and~\ref{thm:4cd}, as well as full details for how to derandomize both algorithms.
%
Charikar, Guruswami, and Wirth~\yrcite{CharikarGuruswamiWirth2005} previously showed that the canonical LP relaxations of cluster editing and cluster deletion can be rounded to produce 4-approximation algorithms for both problems. Theorems~\ref{thm:4ce} and~\ref{thm:4cd} show that the same approximation guarantee is possible using LP relaxations with only a subset of constraints. For many natural graph classes (e.g., sparse graphs), the number of constraints is asymptotically smaller.
The \emph{best} approximation algorithms known for these problems---a 2.06-approximation for cluster editing~\cite{ChawlaMakarychevSchrammEtAl2015} and a 2-approximation for cluster deletion~\cite{Veldt:2018:CCF:3178876.3186110}---still use the canonical LP relaxations. However, these come as a significant increase in computational cost. 



\section{Match-Flip-Pivot Algorithms }
We now present a \emph{combinatorial} approach for obtaining lower bounds and approximation algorithms for cluster editing and cluster deletion. We begin by reviewing combinatorial strategies for lower bounding \textsc{minSTC} and \textsc{minSTC+}, which we round in new ways for clustering problems.

\subsection{Lower Bounds via Maximal Matchings}
\label{sec:stcreductions}
The \textsc{minSTC} problem~\eqref{minstc} on a graph $G = (V,E)$ is equivalent to solving vertex cover on the Gallai graph $\mathcal{G}$ of $G$. The Gallai graph~\cite{le1996gallai} has a node $v_{ij}$ for each edge $(i,j) \in E$, and an edge between two nodes $v_{jk}$ and $v_{ik}$ if $(i,j,k)$ is an open wedge centered at $k$ in $G$. The edges in $\mathcal{G}$ are in one-to-one correspondence with the open wedges of $G$. Placing node $v_{ij}$ in the cover can be viewed as labeling the edge $(i,j) \in E$ as \emph{weak}. Since all edges in $\mathcal{G}$ are adjacent to at least one node in any vertex cover, this means that all open wedges in $G$ will have at least on {weak} edge. 

The \textsc{minSTC+} objective~\eqref{mine} can instead be viewed as a vertex cover problem in a three-uniform hypergraph $\mathcal{H} = (V_\mathcal{H}, E_\mathcal{H})$. This hypergraph has a node $v_{ij} \in V_E$ for every \emph{pair} of distinct nodes $(i,j) \in {V \choose 2}$ in the original graph $G = (V,E)$, and a hyperedge $w_{ijk} = \{v_{ij}, v_{ik}, v_{jk}\} \in E_\mathcal{H}$ whenever $(i,j,k)$ is an open wedge in $G$. We will refer to $\mathcal{H}$ as the \emph{open wedge hypergraph}. 

Sintos and Tsaparas~\yrcite{sintos2014using} showed that a 2-approximation for \textsc{minSTC} can be obtained by applying the 2-approximation for vertex cover~\cite{vazirani2001approximation} to the Gallai graph. A similar 3-approximation for \textsc{minSTC+} is obtained by approximating vertex cover on $\mathcal{H}$~\cite{gruttemeierstrong}. These approximate solutions are obtained by first finding a maximal matching in~$\mathcal{G}$ or $\mathcal{H}$ respectively. For our purposes, a maximal matching in $\mathcal{G}$ corresponds to an edge-disjoint set of open wedges in $G$, and lower bounds the cluster deletion objective. Similarly, a maximal matching in $\mathcal{H}$ is a node-pair disjoint set of open wedges in $G$, and lower bounds cluster editing.

\subsection{Math-Flip-Pivot Algorithm for Cluster Editing}
A vertex cover in the Gallai graph corresponds to a feasible STC+ labeling, and hence is a feasible solution for the \textsc{minSTC+} integer program~\eqref{mine}. 
We now show how to round these approximately optimal solutions to \textsc{minSTC+} to approximate cluster editing. We start with a generic approach that rounds any STC+ labeling $(E',E_W)$ into a solution for cluster editing. This algorithms first \emph{flips} edges in the original graph $G = (V,E)$, meaning that we convert some non-adjacent node pairs into edges $E'$, and we delete edges $E_W$ that were previously in $E$. We then run \textsc{Pivot} on the new graph. The number of mistakes made can be bounded in terms of the number of flipped edges.
\begin{theorem}
	\label{thm:maince}
	Let $(E', E_w)$ be an STC+ labeling for graph $G = (V,E)$. If $\hat{E} = E' \cup (E - E_W)$, applying \textsc{Pivot} to derived graph $\hat{G} = (V, \hat{E})$ returns a cluster editing solution for $G$ with at most $2(|E'| + |E_W|)$ mistakes in expectation.
\end{theorem}
We prove this result in the appendix and also provide a de-randomized version.
We can then use any approximation algorithm for \textsc{minSTC+} to approximate cluster editing.
\begin{corollary}
	\label{corce}
	If $\mathcal{A}$ is an $\alpha$-approximation algorithm for \textsc{minSTC+}, applying the algorithm in Theorem~\ref{thm:maince}
	on an \textsc{STC+} labeling $(E',E_W)$ returned by $\mathcal{A}$ 
	produces a $(2\alpha)$-approximation for cluster editing. If $\mathit{OPT}^+$ and $\mathit{OPT}^{CE}$ are optimal solutions to \textsc{STC+} and cluster editing, then $	\mathit{OPT}^+ \leq \mathit{OPT}^{CE} \leq 2\mathit{OPT}^+$.
\end{corollary}
\begin{algorithm}[tb]
	\caption{$\textsc{MatchFlipPivotCE}(G)$}
	\begin{algorithmic}[5]
		\STATE{\bfseries Input:} Graph $G = (V,E)$ 
		\STATE {\bfseries Output:} Clustering of $G$.
		\STATE \textit{Reduce:} Build open wedge hypergraph $\mathcal{H} = (V_\mathcal{H}, E_\mathcal{H})$ 
		\STATE \textit{Match:} Find maximal matching $\mathcal{M} \subseteq E_\mathcal{H}$
		\STATE \textit{Vertex Cover:} 
		\STATE \begin{center}${C} = \{ v_{ij} \in V_\mathcal{H} \colon v_{ij} \in w \text{ for some  $w \in \mathcal{M}$} \}$ \end{center}
		\STATE \textit{STC+ Labeling}:
		\STATE\begin{center}
			$E' = \{ (i,j) \notin E \colon v_{ij} \in  {C} \}$\\
			$E_W = \{ (i,j) \in E \colon v_{ij} \in {C} \}$
		\end{center}
		\STATE Construct $\hat{G} = (V, \hat{E})$ where $\hat{E} = E' \cup (E - E_W)$
		\STATE Return $\textsc{Pivot}(\hat{G})$
	\end{algorithmic}
	\label{alg:6ce}
\end{algorithm}
Combining Corollary~\ref{corce} with the 3-approximation for \textsc{minSTC+} produces a fast algorithm for cluster editing.
\begin{corollary}
	\label{cor:6ce}
	Algorithm~\ref{alg:6ce} is a randomized 6-approximation for cluster editing.
\end{corollary}
Recall that standard \textsc{Pivot} produces a better {expected} approximation factor of 3 and is also faster, as it requires neither \emph{match} nor \emph{flip} steps. Nevertheless, Algorithm~\ref{alg:6ce} provides new benefits and advantages in both theory and practice. First of all, this algorithm can be derandomized using a completely combinatorial approach (Appendix~\ref{app:det}). Previous approaches for derandomizing \textsc{Pivot} require solving the impractical canonical LP~\cite{vanzuylen2009deterministic}. Our approximation factor of 6 is therefore the best approximation guarantee for complete unweighted correlation clustering among methods that are both combinatorial and deterministic. Furthermore, the \emph{match} step of our algorithm is extremely useful in practice, as it provides explicit lower bounds and a posteriori approximation guarantees that are typically much better than 3. In our experimental results, we find that using our lower bounds in conjunction with standard \textsc{Pivot} yields a very fast method that produces a posteriori approximations that are much better than the 3-approximate a priori guarantee for \textsc{Pivot}.

\subsection{Match-Flip-Pivot Algorithm for Cluster Deletion}
We analogously round an approximate feasible solution to \textsc{minSTC} to approximate cluster deletion. Recall that a feasible solution to \textsc{minSTC} for a graph $G = (V,E)$ is a set of edges $E_W$ to label as weak to ensure all open wedges in $G$ have at least one weak edge. We prove that applying \textsc{Pivot} to $G$ after deleting all of the edges in $E_W$ will produce a feasible cluster deletion solution with a bound on the number of deleted edges. Proofs and derandomized algorithms are given in the appendix.
\begin{algorithm}[tb]
	\caption{$\textsc{MatchFlipPivotCD}(G)$}
	\begin{algorithmic}[5]
		\STATE{\bfseries Input:} Graph $G = (V,E)$ 
		\STATE {\bfseries Output:} Feasible cluster deletion clustering of $G$.
		\STATE \textit{Reduce:} Build Gallai graph $\mathcal{G} = (V_\mathcal{G}, E_\mathcal{G})$ (Section~\ref{sec:stcreductions})
		\STATE \textit{Match:} Find maximal matching $\mathcal{M} \subseteq E_\mathcal{G}$
		\STATE \textit{Cover:} $\mathcal{C} = \{ v_{ij} \in V_\mathcal{G} \colon v_{ij} \in w \text{ for some $w \in \mathcal{M}$} \}$ 
		\STATE \textit{STC Labeling}: $E_W = \{ (i,j) \in E \colon v_{ij} \in  \mathcal{C} \}$
		\STATE Construct graph $\hat{G} = (V, \hat{E})$ where $\hat{E} = (E - E_W)$
		\STATE Return $\textsc{Pivot}(\hat{G})$
	\end{algorithmic}
	\label{alg:mfpcd}
\end{algorithm}
\begin{theorem}
	\label{thm:maincd}
	Let $E_W$ be an STC label set for graph $G = (V,E)$. Applying \textsc{Pivot} to graph $\hat{G} = (V, E-E_W)$ returns a cluster deletion solution for $G$ with at most $2|E_W|$ 
	deleted edges in expectation.
\end{theorem}
We obtain the following corollary on the relationship between \textsc{minSTC} and cluster deletion. 
\begin{corollary}
	\label{corcd}
	If $\mathcal{A}$ is an $\alpha$-approximation algorithm for \textsc{minSTC}, running the algorithm in Theorem~\ref{thm:maincd} 
	with an \textsc{STC} label set returned by $\mathcal{A}$ produces a $(2\alpha)$-approximation for cluster deletion. If $\mathit{OPT}^{STC}$ and $\mathit{OPT}^{CD}$ are optimal solutions to \textsc{minSTC} and cluster deletion, then $\mathit{OPT}^{STC} \leq \mathit{OPT}^{CD} \leq 2\mathit{OPT}^{STC}$.
\end{corollary}

We can also use Theorem~\ref{thm:maincd} to obtain the first combinatorial approximation algorithm for cluster deletion. 
\begin{corollary}
	\label{cor:4cd}
	Algorithm~\ref{alg:mfpcd} is a randomized 4-approximation algorithm for cluster deletion.
\end{corollary}
A derandomized (and still combinatorial) version of the algorithm is provided in Appendix~\ref{app:det}.



\section{Experiments}
In practice, our algorithms are far more scalable than the best LP relaxation algorithms, and can be run on graphs that are orders of magnitude larger, at a very small loss in approximation guarantee. 
We demonstrate this by computing lower bounds and approximate solutions for cluster deletion and cluster editing problems on a range of different types of public graph datasets from the Suitesparse matrix collection~\cite{suitesparse2011davis}, the SNAP graph collection~\cite{snapnets}, and the Facebook100 collection~\cite{traud2012facebook}. We also run experiments on the 2021 PACE Challenge benchmark graphs for cluster editing~\cite{kellerhals2021pace}. 
Cluster editing and cluster deletion were first motivated by applications to clustering biological networks~\cite{Ben-DorShamirYakhini1999,ShamirSharanTsur2004}. These problems can also be viewed as special cases of more general frameworks for community detection in unsigned graphs~\cite{Veldt:2018:CCF:3178876.3186110}, which is why we consider solving them on graphs from a range of different application domains (e.g., social networks, biological networks, email networks, collaboration networks). Our experiments are run on a MacBook Air with 16GB of RAM and an Apple M1 chip. All of our algorithms are implemented in the Julia programming language, and we use Gurobi software to solve linear programming relaxations. 
Code for all of our algorithms and experimental results are available at~\url{github.com/nveldt/FastCC-via-STC}.

\subsection{Cluster Deletion Approximation Algorithms}
For cluster deletion, we run our strong triadic closure LP rounding algorithm (LP-STC, Algorithm~\ref{alg:4cd}), our match-flip-pivot technique (MFP-CD, Algorithm~\ref{alg:mfpcd}), and compare these against solving and rounding the canonical LP relaxation (LP-CD). All previous approximation algorithms for cluster deletion rely on the canonical LP~\cite{CharikarGuruswamiWirth2005,puleo2015correlation,Veldt:2018:CCF:3178876.3186110}. We specifically compare against the rounding procedure that provides the best-case 2-approximation~\cite{Veldt:2018:CCF:3178876.3186110}. All the lower bounds we consider can be rounded with either a deterministic or randomized pivoting procedure on some type of derived graph. The bottleneck in all runtimes is computing the lower bound, so for these experiments we apply randomized \textsc{Pivot} 100 times on each derived graph and take the best result as this is simple, fast, and effective.

Table~\ref{tab:snapshort} shows results on 4 of the larger graphs we consider. In the appendix, we show full results for a wider range of graphs that vary in size. We find overall that \textbf{LP-STC is roughly twice as fast as LP-CD, while matching it in solution quality}. This method always returns the \emph{same lower bound} as LP-CD. In some cases it outputs a solution that is also feasible for the canonical relaxation. In other cases, the solution is not feasible for the canonical relaxation but still has a matching objective score. The differences in rounded solutions for these methods is negligible (they trade off in performance) and is likely due to slight variations in the randomized rounding pivot procedure rather than the lower bound itself. Meanwhile, \textbf{MFP-CD is 2-3 \emph{orders of magnitude} faster than LP-CD, with very minor loss to approximation factor}. A posteriori approximation guarantees are always at or below 2.12, and are usually below 2. The lower bound is only slightly worse than the bounds from the LP methods in all cases. In 21 out of 24 cases, MFP-CD in fact returns a better rounded solution (see appendix).
\begin{table}
	\caption{
		Lower bounds (LB), rounded objective scores (UB), approximation ratios (Ratio) and runtimes (Run) for MFP-CD, LP-STC, and the existing 2-approximation for cluster deletion (LP-CD,~\citet{Veldt:2018:CCF:3178876.3186110}). Larger LB is better; smaller UB is better. The lower bound for LP-STC is guaranteed to be bounded above by the bound for LP-CD. In practice, the former is twice as fast, while outputting a matching lower bound in all cases. MFP-CD is theoretically guaranteed to produce even looser lower bounds, but these are still very close in practice and the method is 2-3 orders of magnitude faster. In many cases, MFP-CD actually produces better rounded solutions (best shown in bold), even if the approximation ratio is slightly higher because of a looser lower bound.
	}
	\label{tab:snapshort}
	\centering
	\scalebox{0.95}{\begin{tabular}{lllll}
			\toprule
			\textbf{Graph} & & MFP-CD & LP-STC & LP-CD \\
			\midrule
			\textsc{EmailEnron} & LB & 84385 & 87861.0 & 87861.0  \\
			& UB & \textbf{169793} & 173936.0 & 174035.0  \\
			$n = 36692$ & Ratio & 2.012 & 1.98 & 1.981 \\
			$m = 183831$ & Run & 0.398 & 243.0 & 391.0 \\
			\midrule
			\textsc{condmat05} & LB & 72428 & 79287.5 & 79287.5  \\
			& UB & \textbf{147826} & 152791.0 & 153446.0  \\
			$n = 36458$ & Ratio & 2.041 & 1.927 & 1.935 \\
			$m = 171734$ & Run & 0.433 & 39.5 & 72.5 \\
			\midrule
			\textsc{caAstroPh} & LB & 87563 & 91188.0 & 91188.0  \\
			& UB & 178278 & 174918.0 & \textbf{174802.0}  \\
			$n = 17903$ & Ratio & 2.036 & 1.918 & 1.917 \\
			$m = 196972$ & Run & 0.367 & 78.4 & 376.0 \\
			\midrule
			\textsc{loc-} & LB & 101924 & 106429.0 & 106429.0  \\
			\textsc{Brightkite}& UB & \textbf{204104} & 211219.0 & 211240.0  \\
			$n = 58228$ & Ratio & 2.003 & 1.985 & 1.985 \\
			$m = 214078$ & Run & 0.632 & 151.0 & 241.0 \\
			\bottomrule
	\end{tabular}}
\end{table} 

\subsection{Cluster Editing Approximation Algorithms}
\begin{table}[t]
	\caption{
		Lower bounds (LB), rounded solution scores (UB), approximation ratios (Ratio) and runtimes (Run) for  MFP-CE, LP-STC+, and the 2.06 approximation for cluster editing (LP-CE,~\citet{ChawlaMakarychevSchrammEtAl2015}). 
		To even run LP-CE, we first solve the STC+ relaxation and iteratively add violated constraints. This often (but not always) makes it possible to run LP-CE almost as quickly as LP-STC+. 
	MFP-CE obtains good approximations in under a second for problems that are so large the LP methods run out of memory (indicated by a dash).
	}
	\label{tab:ccshort}
	\centering
	\scalebox{0.95}{\begin{tabular}{lllll}
			\toprule
			\textbf{Graph} & & MFP-CE & LP-STC+ & LP-CE \\
			\midrule
			\textsc{caHepTh} & LB & 10609  & 11289.8  &11290.5  \\
										& UB &29049 &  21625& 20597  \\
			$n = 8638 $ & Ratio & 2.738 & 1.915 & 1.824  \\
			$m = 24806$ & Run  & 0.049  & 62.6& 562.1  \\
			\midrule
			\textsc{Simmons81} & LB & 16402 &  16490.0 & 16490.0\\
			& UB  & 38856 & 32977& 34391  \\
			$n = 1518 $ & Ratio & 2.369 & 2.0 & 2.086  \\
			$m = 32988$ & Run & 0.0645 & 235.8 & 236.9  \\
			\midrule
			\textsc{caAstroPh} & LB & 86369  & --   &--  \\
			& UB & 225943 & --  & --  \\
			$n = 17903 $ & Ratio & 2.616 &  -- & --  \\
			$m = 196972$ & Run & 0.488  &  -- & --  \\
			\midrule
			\textsc{loc}& LB & 101544 & --  &--  \\
			\textsc{-Brightkite}& UB & 267453 & -- & --  \\
			$n = 58228 $ & Ratio & 2.634 & -- & --  \\
			$m = 214078$ & Run &  0.559  & -- & --  \\
			\bottomrule
	\end{tabular}}
\end{table} 
We run a similar set of experiments for cluster editing (CE), i.e., complete unweighted correlation clustering, with the same overall findings. Table~\ref{tab:ccshort} summarizes results for a few graphs. Our method for rounding the \textsc{minSTC+} LP relaxation (\textsc{LP-STC+}, Algorithm~\ref{alg:4owl}), is noticeably faster than solving and rounding the canonical LP-relaxation (\textsc{LP-CE},~\citet{ChawlaMakarychevSchrammEtAl2015}), and produces nearly identical approximation results. As a bonus, \textsc{LP-STC+} can be used as a first step of a more efficient approach for solving the full canonical relaxation for cluster editing. 
 In fact, we are not even able to solve the canonical relaxation on graphs with a few hundred nodes without using a ``lazy constraints'' method that involves first applying \textsc{LP-STC+} and then iteratively updating constraints (see the appendix for further details). If we tried forming the entire constraint matrix for the canonical LP relaxation, LP-CE would not be able to run on the graphs in Table~\ref{tab:ccshort}. In comparison with LP-based methods, our match-flip-pivot method (MFP-CE, Algorithm~\ref{alg:6ce}) is orders of magnitude faster than the LP methods. \textbf{Even for problems where LP-based methods run out of memory, MFP-CE obtains good results in under a second}. The appendix includes results on more graphs, and provides additional details and results for various alternative rounding schemes. For example, we find that combining the lower bounds from MFP-CE with the clusterings obtained by running standard \textsc{Pivot} leads to significantly improved a posteriori approximation guarantees, with no increase in runtime.

\subsection{Match-Flip-Pivot on Large Graphs}
\label{sec:big}
\begin{figure}
	\centering
	\subfigure[Cluster deletion approx vs.\ cluster editing approx \label{fig:mfpcdvcc} ]
	{\includegraphics[width=.9\linewidth]{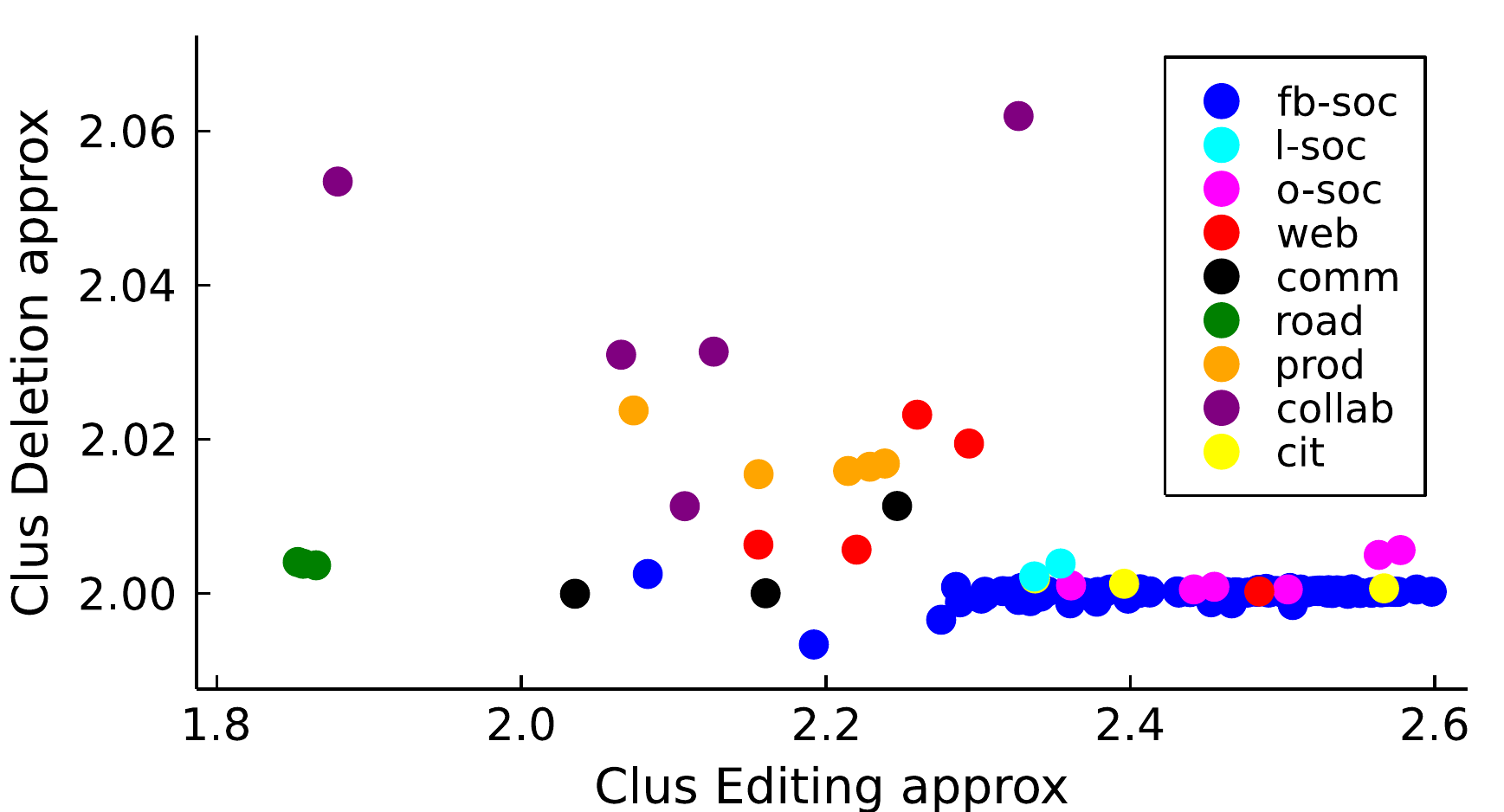}}
	\subfigure[MFP-CD runtimes] 
	{\includegraphics[width=.49\linewidth]{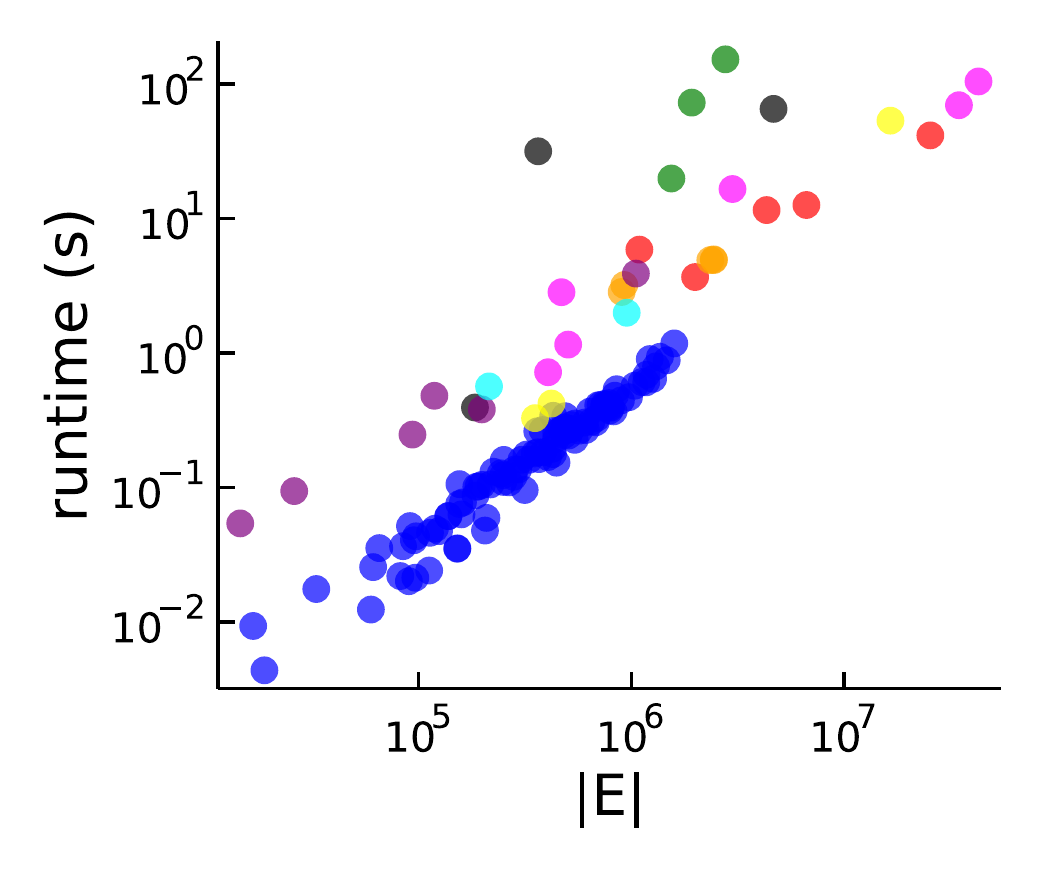}}\hfill
	\subfigure[MFP-CE runtimes] 
	{\includegraphics[width=.49\linewidth]{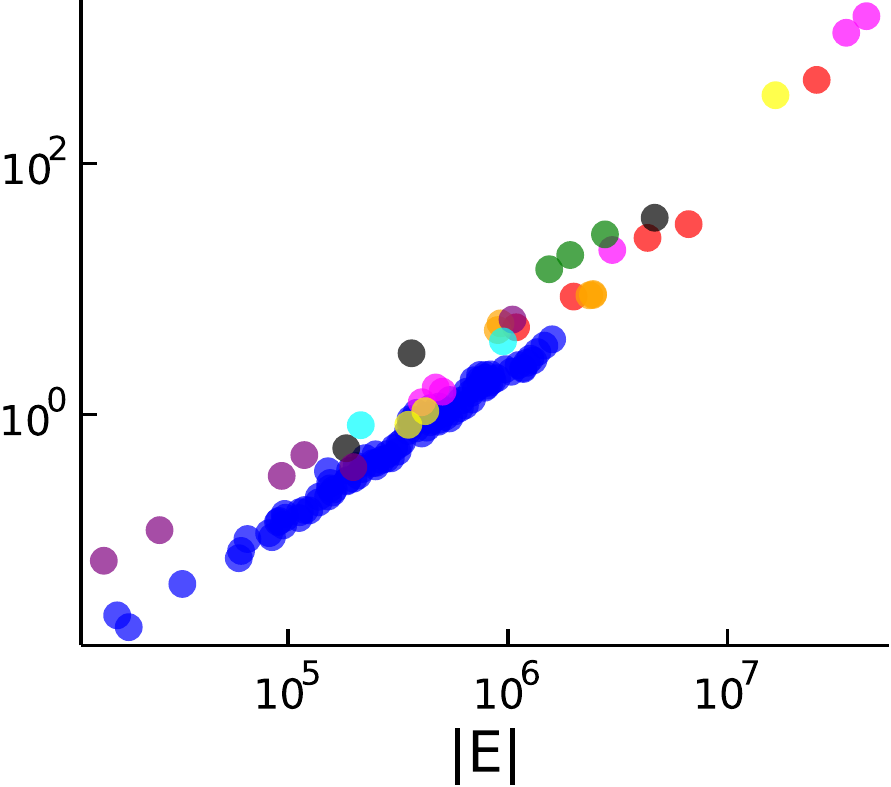}}\hfill
	\vspace{-.5\baselineskip}
	\caption{(a) Our algorithms allow us to observe different patterns in how easy or hard it is to approximate cluster editing and cluster deletion in different graph classes. Runtimes for our match-flip-pivot algorithms for cluster deletion (b) and cluster editing (c) are extremely fast, and scale roughly linearly in term of $|E|$.}
	\label{fig:mfp}
	\vspace{-\baselineskip} 
\end{figure}
Next we test the scalability of our match-flip-pivot techniques on graphs with up to millions of nodes and billions of edges. We run MFP-CE and MFP-CD on all graphs from the Facebook100 dataset~\cite{traud2012facebook}, and on a wide range of SNAP graphs~\cite{snapnets}.  Figure~\ref{fig:mfpcdvcc} reports our a posteriori approximation guarantees for cluster deletion vs.\ our approximations for cluster editing, on graphs from 8 different classes (e.g., Facebook social networks, other social networks, road networks; see appendix for more details). For MFP-CE, we report the approximation factor obtained by combining our lower bounds with the clusterings obtained by standard \textsc{Pivot}, as this is just as fast and tends to produce better results. Our results allow us to highlight and detect interesting patterns that arise for different classes of graphs when it comes to solving these edge modification problems. For example, the three road networks (green) are unique in that cluster editing approximations for these graphs are significantly better than cluster deletion approximations. Meanwhile, the collaboration networks (purple) exhibit worse cluster deletion approximation results than other graphs in general. Finally, social networks are characterized by better cluster deletion approximations, and poorer cluster editing approximations.

These results and observations about graph classes would not be possible if our algorithms were not extremely scalable. Figure~\ref{fig:mfp} shows that our methods scale roughly linearly in terms of the number of edges. 
For most graphs, our methods take a few seconds or less, and only take a few minutes for large graphs with millions of nodes and edges. The largest graph we consider (soc-Livejournal1) has 4.2 million nodes and 4.7 billion edges. We find a 2.01 approximation for cluster deletion in 100 seconds, and a 2.56 approximation for cluster editing in 24 minutes. 
By comparison, standard LP solvers run out of memory for problems that take us less than a second. Specialized solvers for correlation clustering relaxations have also been designed~\cite{ruggles2020parallel,sonthalia2020project,veldt2019metric}. These are more memory efficient than standard black-box optimization methods and can be applied to more general problems, but can be very slow in practice. 
We tried these methods as well but found they were not competitive on the problems we consider. See appendix for details.

\subsection{A Posteriori Guarantees for Fast Heuristics}
\begin{figure}[t]
	\centering
	\subfigure[SNAP CE Ratios] 
	{\includegraphics[width=.49\linewidth]{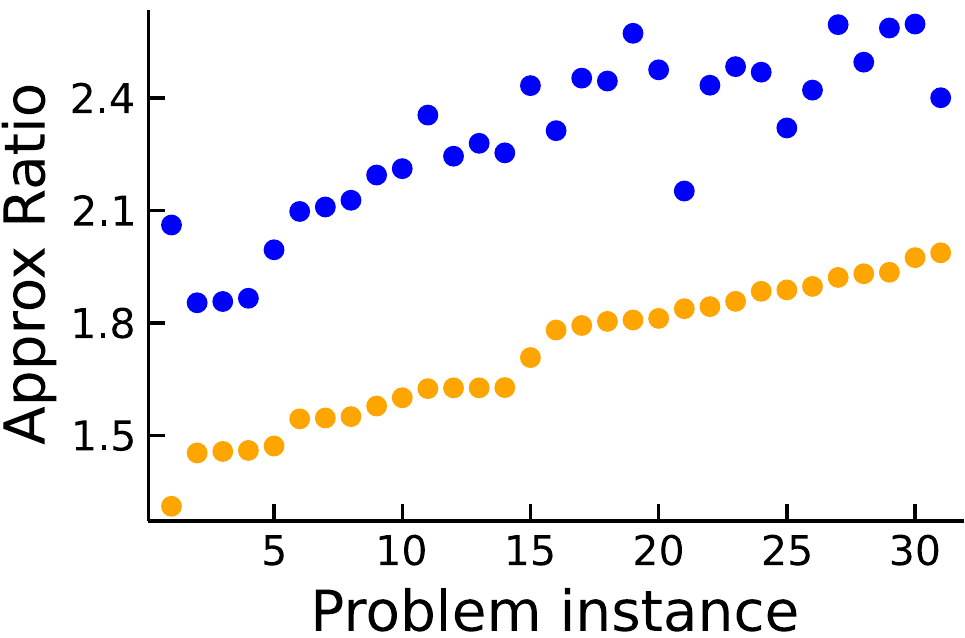}}\hfill
	\subfigure[FB100 CE  Ratios] 
	{\includegraphics[width=.49\linewidth]{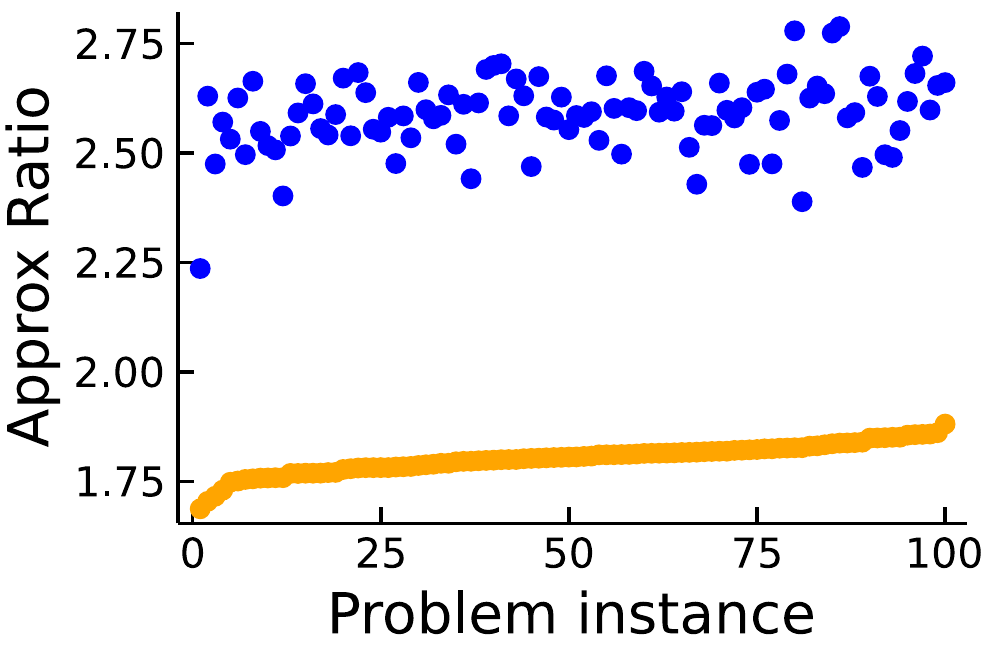}}\hfill
	\subfigure[SNAP CE  Runtimes] 
	{\includegraphics[width=.49\linewidth]{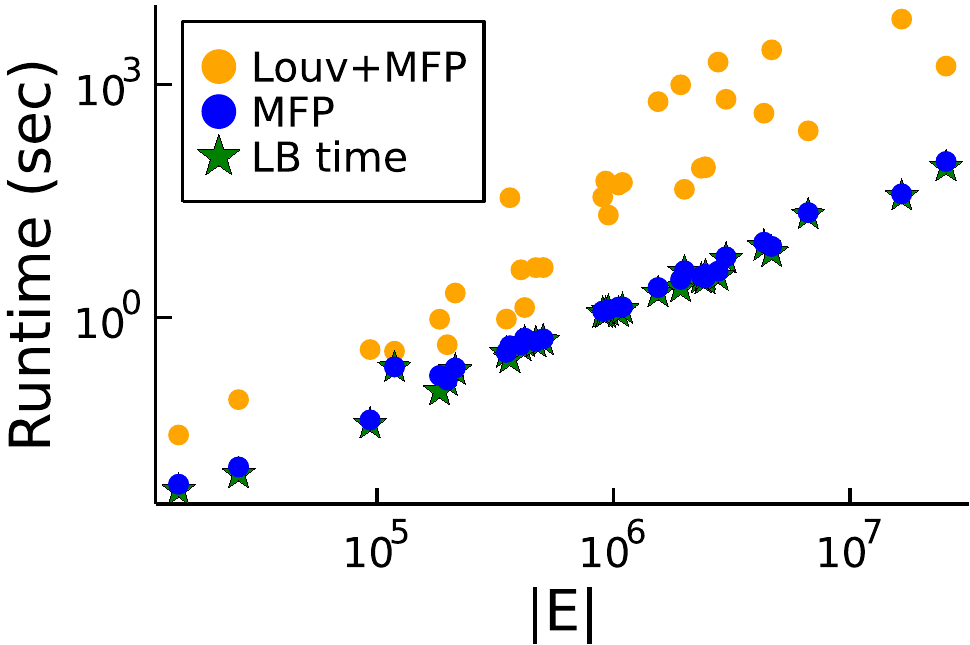}}\hfill
	\subfigure[FB100 CE Runtimes] 
	{\includegraphics[width=.49\linewidth]{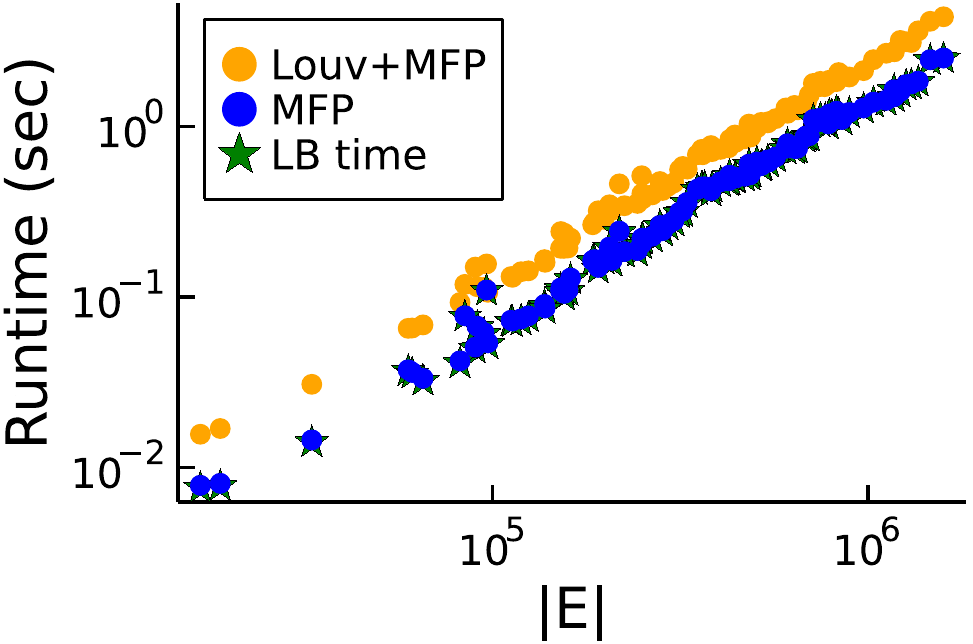}}\hfill
	\vspace{-.5\baselineskip}
	\caption{Approximation ratios and runtimes for MFP-CE, and for combining lower bounds from MFP-CE with Louvain-based heuristics (Louv-MFP), on SNAP and Facebook100 graphs. Runtimes for both methods include the time it takes to obtain the MFP lower bound. We separately show the time it takes to compute this lower bound (LB time). The time it takes MFP-CE to round this lower bound in negligible, but the Louvain approach is slower.}
	\label{fig:louvainsnap}
	\vspace{-\baselineskip} 
\end{figure}

The lower bounds computed by our match-flip-pivot techniques can be used to provide a posteriori approximation guarantees for fast heuristic algorithms that come with no approximation guarantees of their own. To illustrate this, we run a heuristic method for correlation clustering called \textsc{LambdaLouvain}~\cite{Veldt:2018:CCF:3178876.3186110} based on the popular Louvain method for graph clustering~\cite{blondel2008fast}. Figure~\ref{fig:louvainsnap} displays the approximation guarantees we obtain for the cluster editing objective (CE) by combining our MFP lower bounds with \textsc{LambdaLouvain}. Overall this leads to significantly improved a posteriori approximation ratios in comparison with running MFP by itself, at the expense of slower runtimes. Viewed from another perspective, this shows that we can obtain lower bounds to certify that heuristic methods provide approximately optimal solutions, in significantly less time than it takes to actually run these heuristic methods. The appendix provides additional details.
	\vspace{-\baselineskip} 
\subsection{Comparisons on PACE Graphs}
\begin{figure}[t]
	\centering
	\subfigure[PACE Graphs Ratios] 
	{\includegraphics[width=.49\linewidth]{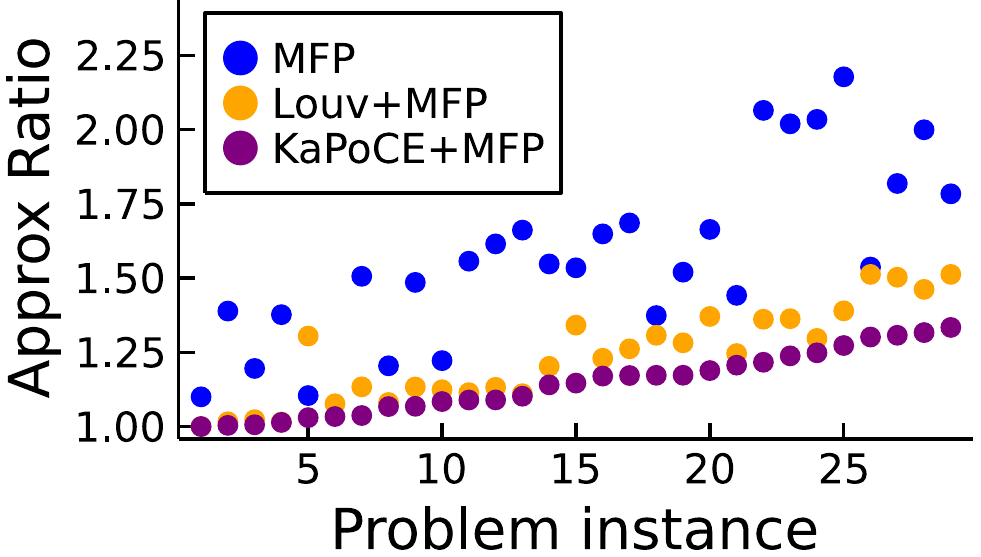}}\hfill
		\subfigure[PACE Graphs Runtimes] 
	{\includegraphics[width=.49\linewidth]{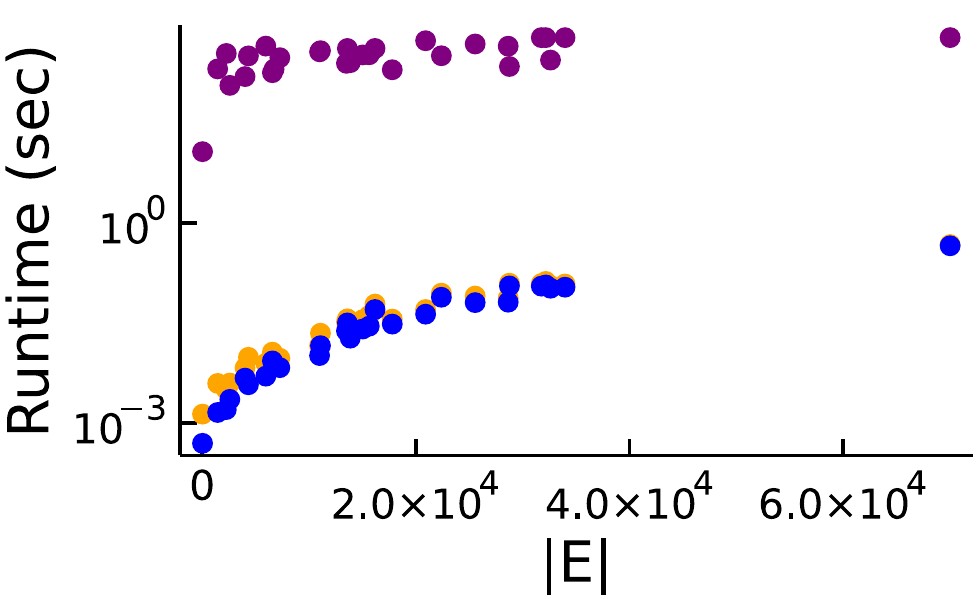}}\hfill
	\vspace{-.5\baselineskip}
	\caption{Comparison against KaPoCE on PACE graphs.}
	\label{fig:pace}
	\vspace{-\baselineskip} 
\end{figure}

Finally, we illustrate the performance of our methods on graphs from the 2021 Parameterized Algorithms and Computational Experiments (PACE) challenge on algorithms for solving the cluster editing objective~\cite{kellerhals2021pace}. We compare against the winning method KaPoCE, which comes with an exact version for finding optimal solutions and a heuristic version with no approximation guarantees. The exact version times out on even on some problems with under 100 nodes, and even the heuristic approach does not scale to large graphs. Figure~\ref{fig:pace} shows results for MFP-CE, compared with results for combining MFP-CE lower bounds with \textsc{LambdaLouvain} and the heuristic KaPoCE algorithm. KaPoCE finds high quality solutions on small graphs, but comes with no approximation guarantees of its own and is orders of magnitude slower than MFP-CE.

\section{Discussion and Open Questions}
We have presented new approximation algorithms for cluster editing and cluster deletion, based on new ways to round lower bounds for related edge labeling problems. We proved that cluster deletion and \textsc{minSTC} are always within a factor of 2; previous work has shown cases where their objectives differ by a factor of $8/7$~\cite{gruttemeier2020relation}. One open question is whether we can tighten these bounds in either direction, or tighten the corresponding bounds between \textsc{minSTC+} and cluster editing. Our research also motivates further work on improved lower bounds and approximation algorithms for correlation clustering that do not rely on LP relaxations. Is it possible to obtain even better approximation guarantees using alternative rounding schemes or simply alternative analyses? One particularly compelling open question is to see whether we can obtain a deterministic combinatorial \textsc{Pivot} method that is a 3-approximation, rather than a 6-approximation, or a combinatorial 2-approximation for cluster deletion. Finally, perhaps the most interesting and meaningful direction for future work is to see whether our fast and practical \text{match-flip-pivot} lower bounds can be used to obtain faster approximation algorithms for weighted variants of correlation clustering whose only approximation algorithms currently rely on LP relaxations~\cite{jafarov2020ccasymmetric,Veldt:2018:CCF:3178876.3186110,puleo2015correlation}.

\bibliography{cc-bib,cd-bib}

\begin{thebibliography}{50}
\providecommand{\natexlab}[1]{#1}
\providecommand{\url}[1]{\texttt{#1}}
\expandafter\ifx\csname urlstyle\endcsname\relax
  \providecommand{\doi}[1]{doi: #1}\else
  \providecommand{\doi}{doi: \begingroup \urlstyle{rm}\Url}\fi

\bibitem[Ailon et~al.(2008)Ailon, Charikar, and
  Newman]{AilonCharikarNewman2008}
Ailon, N., Charikar, M., and Newman, A.
\newblock Aggregating inconsistent information: ranking and clustering.
\newblock \emph{Journal of the ACM (JACM)}, 55\penalty0 (5):\penalty0 23, 2008.

\bibitem[Bansal et~al.(2004)Bansal, Blum, and Chawla]{BansalBlumChawla2004}
Bansal, N., Blum, A., and Chawla, S.
\newblock Correlation clustering.
\newblock \emph{Machine Learning}, 56:\penalty0 89--113, 2004.

\bibitem[Beier et~al.(2015)Beier, Hamprecht, and Kappes]{Beier_2015_CVPR}
Beier, T., Hamprecht, F.~A., and Kappes, J.~H.
\newblock Fusion moves for correlation clustering.
\newblock In \emph{Proceedings of the IEEE Conference on Computer Vision and
  Pattern Recognition (CVPR)}, June 2015.

\bibitem[Ben-Dor et~al.(1999)Ben-Dor, Shamir, and
  Yakhini]{Ben-DorShamirYakhini1999}
Ben-Dor, A., Shamir, R., and Yakhini, Z.
\newblock Clustering gene expression patterns.
\newblock \emph{Journal of computational biology}, 6\penalty0 (3-4):\penalty0
  281--297, 1999.

\bibitem[Bhattacharya \& De(2008)Bhattacharya and De]{bhattacharya2008divisive}
Bhattacharya, A. and De, R.~K.
\newblock Divisive correlation clustering algorithm (dcca) for grouping of
  genes: detecting varying patterns in expression profiles.
\newblock \emph{bioinformatics}, 24\penalty0 (11):\penalty0 1359--1366, 2008.

\bibitem[Bhattacharya \& De(2010)Bhattacharya and De]{bhattacharya2010average}
Bhattacharya, A. and De, R.~K.
\newblock Average correlation clustering algorithm (acca) for grouping of
  co-regulated genes with similar pattern of variation in their expression
  values.
\newblock \emph{Journal of Biomedical Informatics}, 43\penalty0 (4):\penalty0
  560--568, 2010.
\newblock ISSN 1532-0464.
\newblock \doi{https://doi.org/10.1016/j.jbi.2010.02.001}.
\newblock URL
  \url{https://www.sciencedirect.com/science/article/pii/S1532046410000158}.

\bibitem[Blondel et~al.(2008)Blondel, Guillaume, Lambiotte, and
  Lefebvre]{blondel2008fast}
Blondel, V.~D., Guillaume, J.-L., Lambiotte, R., and Lefebvre, E.
\newblock Fast unfolding of communities in large networks.
\newblock \emph{Journal of statistical mechanics: theory and experiment},
  2008\penalty0 (10):\penalty0 P10008, 2008.

\bibitem[Bonchi et~al.(2022)Bonchi, Garc{\'\i}a-Soriano, and
  Gullo]{bonchi2022correlation}
Bonchi, F., Garc{\'\i}a-Soriano, D., and Gullo, F.
\newblock \emph{Correlation Clustering: Morgan \& Claypool Publishers}.
\newblock Morgan \& Claypool Publishers, 2022.

\bibitem[Bun et~al.(2021)Bun, Elias, and Kulkarni]{bun2021differentially}
Bun, M., Elias, M., and Kulkarni, J.
\newblock Differentially private correlation clustering.
\newblock In Meila, M. and Zhang, T. (eds.), \emph{Proceedings of the 38th
  International Conference on Machine Learning}, volume 139 of
  \emph{Proceedings of Machine Learning Research}, pp.\  1136--1146. PMLR,
  18--24 Jul 2021.
\newblock URL \url{https://proceedings.mlr.press/v139/bun21a.html}.

\bibitem[Charikar et~al.(2005)Charikar, Guruswami, and
  Wirth]{CharikarGuruswamiWirth2005}
Charikar, M., Guruswami, V., and Wirth, A.
\newblock Clustering with qualitative information.
\newblock \emph{Journal of Computer and System Sciences}, 71\penalty0
  (3):\penalty0 360 -- 383, 2005.
\newblock ISSN 0022-0000.
\newblock \doi{http://dx.doi.org/10.1016/j.jcss.2004.10.012}.
\newblock URL
  \url{http://www.sciencedirect.com/science/article/pii/S0022000004001424}.
\newblock Learning Theory 2003.

\bibitem[Chawla et~al.(2015)Chawla, Makarychev, Schramm, and
  Yaroslavtsev]{ChawlaMakarychevSchrammEtAl2015}
Chawla, S., Makarychev, K., Schramm, T., and Yaroslavtsev, G.
\newblock Near optimal lp rounding algorithm for correlation clustering on
  complete and complete k-partite graphs.
\newblock In \emph{Proceedings of the Forty-Seventh Annual ACM on Symposium on
  Theory of Computing}, pp.\  219--228. ACM, 2015.

\bibitem[Chen et~al.(2012)Chen, Sanghavi, and Xu]{chen2012clustering}
Chen, Y., Sanghavi, S., and Xu, H.
\newblock Clustering sparse graphs.
\newblock In \emph{Advances in Neural Information Processing Systems},
  volume~25 of \emph{NIPS '12}. Curran Associates, Inc., 2012.
\newblock URL
  \url{https://proceedings.neurips.cc/paper/2012/file/1e6e0a04d20f50967c64dac2d639a577-Paper.pdf}.

\bibitem[Chierichetti et~al.(2014)Chierichetti, Dalvi, and
  Kumar]{chierichetti2014correlation}
Chierichetti, F., Dalvi, N., and Kumar, R.
\newblock Correlation clustering in mapreduce.
\newblock In \emph{Proceedings of the 20th ACM SIGKDD international conference
  on Knowledge discovery and data mining}, pp.\  641--650, 2014.

\bibitem[Cohen-Addad et~al.(2021)Cohen-Addad, Lattanzi, Mitrovi\'{c},
  Norouzi-Fard, Parotsidis, and Tarnawski]{cohen-addad2021correlation}
Cohen-Addad, V., Lattanzi, S., Mitrovi\'{c}, S., Norouzi-Fard, A., Parotsidis,
  N., and Tarnawski, J.
\newblock Correlation clustering in constant many parallel rounds.
\newblock In \emph{Proceedings of the 38th International Conference on Machine
  Learning}, ICML '21. PMLR, July 2021.

\bibitem[Davis \& Hu(2011)Davis and Hu]{suitesparse2011davis}
Davis, T.~A. and Hu, Y.
\newblock The university of florida sparse matrix collection.
\newblock \emph{ACM Trans. Math. Softw.}, 38\penalty0 (1), dec 2011.
\newblock ISSN 0098-3500.
\newblock \doi{10.1145/2049662.2049663}.
\newblock URL \url{https://doi.org/10.1145/2049662.2049663}.

\bibitem[Demaine et~al.(2006)Demaine, Emanuel, Fiat, and
  Immorlica]{DemaineEmanuelFiatEtAl2006}
Demaine, E.~D., Emanuel, D., Fiat, A., and Immorlica, N.
\newblock Correlation clustering in general weighted graphs.
\newblock \emph{Theoretical Computer Science}, 361\penalty0 (2):\penalty0 172
  -- 187, 2006.
\newblock ISSN 0304-3975.
\newblock \doi{https://doi.org/10.1016/j.tcs.2006.05.008}.
\newblock URL
  \url{http://www.sciencedirect.com/science/article/pii/S0304397506003227}.
\newblock Approximation and Online Algorithms.

\bibitem[Easley et~al.(2010)Easley, Kleinberg, et~al.]{easley2010networks}
Easley, D., Kleinberg, J., et~al.
\newblock \emph{Networks, crowds, and markets}, volume~8.
\newblock Cambridge university press Cambridge, 2010.

\bibitem[Gleich et~al.(2018)Gleich, Veldt, and Wirth]{gleich2018ccgen}
Gleich, D.~F., Veldt, N., and Wirth, A.
\newblock {Correlation Clustering Generalized}.
\newblock In \emph{29th International Symposium on Algorithms and Computation},
  volume 123 of \emph{ISAAC 2018}, pp.\  44:1--44:13, Dagstuhl, Germany, 2018.
  Schloss Dagstuhl--Leibniz-Zentrum fuer Informatik.
\newblock ISBN 978-3-95977-094-1.
\newblock \doi{10.4230/LIPIcs.ISAAC.2018.44}.
\newblock URL \url{http://drops.dagstuhl.de/opus/volltexte/2018/9992}.

\bibitem[Granovetter(1973)]{granovetter1973strength}
Granovetter, M.~S.
\newblock The strength of weak ties.
\newblock \emph{American journal of sociology}, 78\penalty0 (6):\penalty0
  1360--1380, 1973.

\bibitem[Gr{\"u}ttemeier \& Komusiewicz(2020)Gr{\"u}ttemeier and
  Komusiewicz]{gruttemeier2020relation}
Gr{\"u}ttemeier, N. and Komusiewicz, C.
\newblock On the relation of strong triadic closure and cluster deletion.
\newblock \emph{Algorithmica}, 82\penalty0 (4):\penalty0 853--880, 2020.
\newblock \doi{https://doi.org/10.1007/s00453-019-00617-1}.

\bibitem[Gr{\"u}ttemeier \& Morawietz(2020)Gr{\"u}ttemeier and
  Morawietz]{gruttemeierstrong}
Gr{\"u}ttemeier, N. and Morawietz, N.
\newblock On strong triadic closure with edge insertion.
\newblock \emph{Technical report}, 2020.

\bibitem[Hou et~al.(2016)Hou, Emad, Puleo, Ma, and Milenkovic]{hou2016cancercc}
Hou, J.~P., Emad, A., Puleo, G.~J., Ma, J., and Milenkovic, O.
\newblock A new correlation clustering method for cancer mutation analysis.
\newblock \emph{Bioinformatics}, 32\penalty0 (24):\penalty0 3717--3728, 2016.
\newblock \doi{10.1093/bioinformatics/btw546}.
\newblock URL \url{http://dx.doi.org/10.1093/bioinformatics/btw546}.

\bibitem[Jafarov et~al.(2020)Jafarov, Kalhan, Makarychev, and
  Makarychev]{jafarov2020ccasymmetric}
Jafarov, J., Kalhan, S., Makarychev, K., and Makarychev, Y.
\newblock Correlation clustering with asymmetric classification errors.
\newblock In \emph{Proceedings of the 37th International Conference on Machine
  Learning}, ICML '20, pp.\  4641--4650. PMLR, 13--18 Jul 2020.
\newblock URL \url{https://proceedings.mlr.press/v119/jafarov20a.html}.

\bibitem[Jafarov et~al.(2021)Jafarov, Kalhan, Makarychev, and
  Makarychev]{jafarov2021local}
Jafarov, J., Kalhan, S., Makarychev, K., and Makarychev, Y.
\newblock Local correlation clustering with asymmetric classification errors.
\newblock In \emph{Proceedings of the 36th International Conference on Machine
  Learning}, ICML '21, pp.\  4677--4686. PMLR, 2021.

\bibitem[Kellerhals et~al.(2021)Kellerhals, Koana, Nichterlein, and
  Zschoche]{kellerhals2021pace}
Kellerhals, L., Koana, T., Nichterlein, A., and Zschoche, P.
\newblock The pace 2021 parameterized algorithms and computational experiments
  challenge: Cluster editing.
\newblock In \emph{16th International Symposium on Parameterized and Exact
  Computation (IPEC 2021)}. Schloss Dagstuhl-Leibniz-Zentrum f{\"u}r
  Informatik, 2021.

\bibitem[Kim et~al.(2011)Kim, Nowozin, Kohli, and Yoo]{kim2011highcc}
Kim, S., Nowozin, S., Kohli, P., and Yoo, C.~D.
\newblock Higher-order correlation clustering for image segmentation.
\newblock In Shawe-Taylor, J., Zemel, R.~S., Bartlett, P.~L., Pereira, F., and
  Weinberger, K.~Q. (eds.), \emph{Advances in Neural Information Processing
  Systems 24}, pp.\  1530--1538. Curran Associates, Inc., 2011.

\bibitem[Konstantinidis et~al.(2018)Konstantinidis, Nikolopoulos, and
  Papadopoulos]{konstantinidis2018strong}
Konstantinidis, A.~L., Nikolopoulos, S.~D., and Papadopoulos, C.
\newblock Strong triadic closure in cographs and graphs of low maximum degree.
\newblock \emph{Theoretical Computer Science}, 740:\penalty0 76--84, 2018.

\bibitem[Lange et~al.(2018)Lange, Karrenbauer, and Andres]{lange2018partial}
Lange, J.-H., Karrenbauer, A., and Andres, B.
\newblock Partial optimality and fast lower bounds for weighted correlation
  clustering.
\newblock In \emph{Proceedings of the 35th International Conference on Machine
  Learning}, pp.\  2892--2901. PMLR, 2018.

\bibitem[Le(1996)]{le1996gallai}
Le, V.~B.
\newblock Gallai graphs and anti-gallai graphs.
\newblock \emph{Discrete Mathematics}, 159\penalty0 (1-3):\penalty0 179--189,
  1996.

\bibitem[Leskovec \& Krevl(2014)Leskovec and Krevl]{snapnets}
Leskovec, J. and Krevl, A.
\newblock {SNAP Datasets}: {Stanford} large network dataset collection.
\newblock \url{http://snap.stanford.edu/data}, June 2014.

\bibitem[Levinkov et~al.(2017)Levinkov, Kirillov, and
  Andres]{levinkov2017comparative}
Levinkov, E., Kirillov, A., and Andres, B.
\newblock A comparative study of local search algorithms for correlation
  clustering.
\newblock In \emph{German Conference on Pattern Recognition}, pp.\  103--114.
  Springer, 2017.

\bibitem[{Li} et~al.(2019){Li}, {Puleo}, and {Milenkovic}]{li2019motif}
{Li}, P., {Puleo}, G.~J., and {Milenkovic}, O.
\newblock Motif and hypergraph correlation clustering.
\newblock \emph{IEEE Transactions on Information Theory}, pp.\  1--1, 2019.
\newblock ISSN 1557-9654.
\newblock \doi{10.1109/TIT.2019.2940246}.

\bibitem[Pan et~al.(2015)Pan, Papailiopoulos, Oymak, Recht, Ramchandran, and
  Jordan]{xinghao2015parallel}
Pan, X., Papailiopoulos, D., Oymak, S., Recht, B., Ramchandran, K., and Jordan,
  M.~I.
\newblock Parallel correlation clustering on big graphs.
\newblock In \emph{Proceedings of the 28th International Conference on Neural
  Information Processing Systems}, NIPS '15, pp.\  82--90, Cambridge, MA, USA,
  2015. MIT Press.

\bibitem[Puleo \& Milenkovic(2015)Puleo and Milenkovic]{puleo2015correlation}
Puleo, G. and Milenkovic, O.
\newblock Correlation clustering with constrained cluster sizes and extended
  weights bounds.
\newblock \emph{SIAM Journal on Optimization}, 25\penalty0 (3):\penalty0
  1857--1872, 2015.
\newblock \doi{10.1137/140994198}.
\newblock URL \url{https://doi.org/10.1137/140994198}.

\bibitem[Puleo \& Milenkovic(2018)Puleo and Milenkovic]{puleo2018correlation}
Puleo, G.~J. and Milenkovic, O.
\newblock Correlation clustering and biclustering with locally bounded errors.
\newblock \emph{IEEE Transactions on Information Theory}, 64\penalty0
  (6):\penalty0 4105--4119, June 2018.
\newblock ISSN 0018-9448.
\newblock \doi{10.1109/TIT.2018.2819696}.

\bibitem[Ruggles et~al.(2020)Ruggles, Veldt, and Gleich]{ruggles2020parallel}
Ruggles, C., Veldt, N., and Gleich, D.~F.
\newblock A parallel projection method for metric constrained optimization.
\newblock In \emph{Proceedings of the SIAM Workshop on Combinatorial Scientific
  Computing (CSC)}, pp.\  43--53, 2020.
\newblock \doi{10.1137/1.9781611976229.5}.
\newblock URL \url{https://epubs.siam.org/doi/abs/10.1137/1.9781611976229.5}.

\bibitem[Shamir et~al.(2004)Shamir, Sharan, and Tsur]{ShamirSharanTsur2004}
Shamir, R., Sharan, R., and Tsur, D.
\newblock Cluster graph modification problems.
\newblock \emph{Discrete Applied Mathematics}, 144:\penalty0 173--182, 2004.

\bibitem[Shi et~al.(2021)Shi, Dhulipala, Eisenstat, {\L}{\cedilla{a}}cki, and
  Mirrokni]{shi2021scalable}
Shi, J., Dhulipala, L., Eisenstat, D., {\L}{\cedilla{a}}cki, J., and Mirrokni,
  V.
\newblock Scalable community detection via parallel correlation clustering.
\newblock In \emph{VLDB}, 2021.

\bibitem[Sintos \& Tsaparas(2014)Sintos and Tsaparas]{sintos2014using}
Sintos, S. and Tsaparas, P.
\newblock Using strong triadic closure to characterize ties in social networks.
\newblock In \emph{Proceedings of the 20th ACM SIGKDD international conference
  on Knowledge discovery and data mining}, KDD '14, pp.\  1466--1475, 2014.
\newblock URL \url{https://doi.org/10.1145/2623330.2623664}.

\bibitem[Sonthalia \& Gilbert(2020)Sonthalia and Gilbert]{sonthalia2020project}
Sonthalia, R. and Gilbert, A.~C.
\newblock Project and forget: Solving large-scale metric constrained problems.
\newblock \emph{arXiv preprint arXiv:2005.03853}, 2020.

\bibitem[Swoboda \& Andres(2017)Swoboda and Andres]{swoboda2017message}
Swoboda, P. and Andres, B.
\newblock A message passing algorithm for the minimum cost multicut problem.
\newblock In \emph{Proceedings of the 2017 IEEE Conference on Computer Vision
  and Pattern Recognition}, CVPR 2017, pp.\  1617--1626. IEEE, 2017.

\bibitem[Traud et~al.(2012)Traud, Mucha, and Porter]{traud2012facebook}
Traud, A.~L., Mucha, P.~J., and Porter, M.~A.
\newblock {Social structure of Facebook networks}.
\newblock \emph{Physica A: Statistical Mechanics and its Applications},
  391\penalty0 (16):\penalty0 4165--4180, 2012.

\bibitem[Van~Gael \& Zhu(2007)Van~Gael and Zhu]{van2007correlation}
Van~Gael, J. and Zhu, X.
\newblock Correlation clustering for crosslingual link detection.
\newblock In \emph{Proceedings of the 20th International Joint Conference on
  Artifical Intelligence}, IJCAI 2007, pp.\  1744--1749, San Francisco, CA,
  USA, 2007. Morgan Kaufmann Publishers Inc.
\newblock URL \url{http://dl.acm.org/citation.cfm?id=1625275.1625558}.

\bibitem[van Zuylen \& Williamson(2009)van Zuylen and
  Williamson]{vanzuylen2009deterministic}
van Zuylen, A. and Williamson, D.~P.
\newblock Deterministic pivoting algorithms for constrained ranking and
  clustering problems.
\newblock \emph{Mathematics of Operations Research}, 34\penalty0 (3):\penalty0
  594--620, 2009.
\newblock ISSN 0364765X, 15265471.
\newblock URL \url{http://www.jstor.org/stable/40538434}.

\bibitem[Vazirani(2001)]{vazirani2001approximation}
Vazirani, V.~V.
\newblock \emph{Approximation algorithms}, volume~1.
\newblock Springer, 2001.

\bibitem[Veldt et~al.(2018)Veldt, Gleich, and
  Wirth]{Veldt:2018:CCF:3178876.3186110}
Veldt, N., Gleich, D.~F., and Wirth, A.
\newblock A correlation clustering framework for community detection.
\newblock In \emph{Proceedings of the 2018 World Wide Web Conference}, WWW '18,
  pp.\  439--448, Republic and Canton of Geneva, Switzerland, 2018.
  International World Wide Web Conferences Steering Committee.
\newblock ISBN 978-1-4503-5639-8.
\newblock \doi{10.1145/3178876.3186110}.
\newblock URL \url{https://doi.org/10.1145/3178876.3186110}.

\bibitem[Veldt et~al.(2019)Veldt, Gleich, Wirth, and
  Saunderson]{veldt2019metric}
Veldt, N., Gleich, D.~F., Wirth, A., and Saunderson, J.
\newblock Metric-constrained optimization for graph clustering algorithms.
\newblock \emph{SIAM Journal on Mathematics of Data Science}, 1\penalty0
  (2):\penalty0 333--355, 2019.
\newblock \doi{10.1137/18M1217152}.
\newblock URL \url{https://doi.org/10.1137/18M1217152}.

\bibitem[Veldt et~al.(2020)Veldt, Wirth, and Gleich]{veldt2020parameterized}
Veldt, N., Wirth, A., and Gleich, D.~F.
\newblock Parameterized correlation clustering in hypergraphs and bipartite
  graphs.
\newblock In \emph{Proceedings of the 26th ACM SIGKDD International Conference
  on Knowledge Discovery \& Data Mining}, pp.\  1868--1876, 2020.

\bibitem[Wang et~al.(2013)Wang, Xu, Chen, and Wang]{wang2013scalable}
Wang, Y., Xu, L., Chen, Y., and Wang, H.
\newblock A scalable approach for general correlation clustering.
\newblock In Motoda, H., Wu, Z., Cao, L., Zaiane, O., Yao, M., and Wang, W.
  (eds.), \emph{Advanced Data Mining and Applications}, pp.\  13--24, Berlin,
  Heidelberg, 2013. Springer Berlin Heidelberg.
\newblock ISBN 978-3-642-53917-6.

\bibitem[Yarkony et~al.(2012)Yarkony, Ihler, and Fowlkes]{yarkony2012fast}
Yarkony, J., Ihler, A., and Fowlkes, C.~C.
\newblock Fast planar correlation clustering for image segmentation.
\newblock In \emph{European Conference on Computer Vision}, ECCV 2012, pp.\
  568--581, Berlin, Heidelberg, 2012. Springer Berlin Heidelberg.
\newblock ISBN 978-3-642-33783-3.

\end{thebibliography}
\bibliographystyle{icml2022}

\newpage
\appendix
\onecolumn
\section{Proofs for Approximation Algorithms}

\subsection{Proof of Theorem~\ref{thm:4ce}}
\begin{proof}
	We must prove that Theorem~\ref{thm:3pt1} is satisfied with $\alpha = 4$ for an appropriate choice of correlation clustering weights and budgets. For cluster editing, the weights are given by
	\begin{equation}
		\label{4owij}
		(w_{ij}^+, w_{ij}^-) = \begin{cases}
			(1,0) & \text{ if $(i,j) \in E$ }\\
			(0,1) & \text{ if $(i,j) \notin E$,}
		\end{cases}
	\end{equation}
	and the budgets defined by the LP relaxation are given by
	\begin{equation}
		\label{4ocij}
		b_{ij} = \begin{cases}
			x_{ij} & \text{ if $(i,j) \in E$ }\\
			1-x_{ij} & \text{ if $(i,j) \notin E$.}
		\end{cases}
	\end{equation}
	Considering the way $\hat{G}$ is constructed in Algorithm~\ref{alg:4owl}, the conditions in Theorem~\ref{thm:3pt1} translate to the following:
	\begin{enumerate}
		\item If $x_{ij} < 1/2$, we have $w_{ij}^- \leq 4 b_{ij}$, and if $x_{ij} \geq 1/2$, then $w_{ij}^+ \leq 4b_{ij}$.
		\item If $x_{ij} < 1/2$ and $x_{jk} < 1/2$ but $x_{ik} \geq 1/2$, then 
		\begin{equation}
			\label{main4owl}
			w_{ij}^+ + w_{jk}^+ + w_{ik}^- \leq 4(b_{ij} + b_{jk} + b_{ik}).
		\end{equation}
	\end{enumerate}
	Condition 1 is straightforward to check by considering the definitions of edge weights~\eqref{4owij} and budgets~\eqref{4ocij}. We can prove the second condition by case analysis, considering separately whether each node pair $(i,j)$, $(i,k)$, and $(j,k)$ is an edge or not in original graph $G = (V,E)$. Regardless of the case, we have the following bounds, based on the assumption that $(i,j,k)$ is an open wedge centered at $j$ in $\hat{G}$:
	\begin{equation}
		\label{xijbounds4owl}
		1 - x_{ij} > 1/2, \hspace{1cm} 1-x_{jk} > 1/2, \hspace{1cm} x_{ik} \geq 1/2.
	\end{equation}
	
	We summarize all of the cases in succinct tabular format, where we state whether each edge is in $E$ or not, and then give lower bounds on the right hand side of inequality~\eqref{main4owl} to show it is greater than the left hand side in each case. We have ordered cases so that moving from one row to the next changes the edge status of only one node pair at a time, making it easy to quickly see changes in the left and right hand sides of the inequality~\eqref{main4owl} for the corresponding budgets and weights. Several of the bounds we list for the right hand side of~\eqref{main4owl} could be tightened further, but this would not lead to an improved overall approximation guarantee.
	\begin{center}
		\begin{tabular}{cccllc}
			\toprule
			\multicolumn{3}{c}{Is the edge in $E$?} &Right side of~\eqref{main4owl}& Left side of~\eqref{main4owl}& Explanation Note \\
			$(i,j)$ & $(j,k)$ &$(i,k)$ & $4(b_{ij} + b_{jk} + b_{ik})$ & $w_{ij}^+ + w_{jk}^+ + w_{ik}^-$ & \\
			\midrule
			Y&Y&Y &  $4(x_{ij} + x_{jk} + x_{ik}) \geq 4 x_{ik} \geq 2$& 2 = 1+ 1+ 0 & $x_{ik} \geq 1/2$\\
			\midrule
			Y&Y&N &  $4(x_{ij} + x_{jk} + 1-x_{ik}) \geq 4$ & 3 = 1+ 1+ 1 & $x_{ij} + x_{jk} -x_{ik} \geq 0$\\ &&&&&(LP constraint) \\
			\midrule
			Y&N & N& $4(x_{ij} + 1-x_{jk} + 1-x_{ik})> 2 $ & 2 = 1 + 0 + 1 & $1-x_{jk} > 1/2$ \\
			\midrule
			Y& N & Y & $4(x_{ij} + 1-x_{jk} + x_{ik}) \geq 2$ & 1 = 1 + 0 + 0 & $x_{ik}\geq 1/2$ \\
			\midrule
			N& N & Y & $4(1-x_{ij} + 1-x_{jk} + x_{ik})\geq 0$ & 0 = 0 + 0 + 0 & zero left side \\
			\midrule 
			N& Y & Y & $4(1-x_{ij} + x_{jk} + x_{ik}) \geq 2$ & 1 = 0 + 1 + 0 &  $x_{ik}\geq 1/2$  \\
			\midrule 
			N& Y & N & $4(1-x_{ij} + x_{jk} + 1-x_{ik}) > 2$ & 2 = 0 + 1 + 1 &  $1-x_{ij}>1/2$  \\
			\midrule 
			N& N & N & $4(1-x_{ij} + 1-x_{jk} + 1-x_{ik}) > 2$ & 1 = 0 + 0 + 1 &  $1-x_{ij}>1/2$  \\
			\bottomrule
		\end{tabular}
	\end{center}
	
\end{proof}

\subsection{Proof of Theorem~\ref{thm:4cd}}
\begin{proof}
	First of all, note that applying \textsc{Pivot} to the derived graph $\hat{G} = (V,\hat{E})$, using any order of pivot choices, will produce a feasible instance for cluster deletion. To see why, observe that if $k$ is the pivot node and $i$ and $j$ are two of its neighbors in $\hat{G}$, then $z_{ki} < 1/2$ and $z_{kj} < 1/2$, which implies that $(i,j) \in E$. If $(i,j)$ were not an edge, then $(i,j,k)$ would be an open wedge and the LP relaxation would include the constraint $z_{ki} + z_{kj} \geq 1$. It remains to check that Theorem~\ref{thm:3pt1} is satisfied with $\alpha = 4$, for the right choice of budgets and weights. The weights in this case are 
	\begin{equation}
	\label{weightscd}
	(w_{ij}^+, w_{ij}^-) = \begin{cases}
	(1,0) & \text{ if $(i,j) \in E$ }\\
	(0,\infty) & \text{ if $(i,j) \notin E$,}
	\end{cases}
	\end{equation}
	since we are solving cluster deletion. We set our budgets to be the contributions to the LP objective: $b_{ij} = z_{ij}$ if $(i,j) \in E$, and $b_{ij} = 0$ otherwise.
	These conditions we need to satisfy are:
	\begin{enumerate}
		\item If $(i,j) \in \hat{E}$, we have $w_{ij}^- \leq 4 b_{ij}$, and if $(i,j) \notin \hat{E} $, then $w_{ij}^+ \leq 4b_{ij}$.
		\item If $(i,j) \in \hat{E}$ and $(j,k) \in \hat{E}$ and $(i,k) \notin \hat{E}$, then 
		\begin{equation}
		\label{main4cd}
		w_{ij}^+ + w_{jk}^+ + w_{ik}^- \leq 4(b_{ij} + b_{jk} + b_{ik}).
		\end{equation}
	\end{enumerate}
	\textit{Checking condition 1.} Observe that $(i,j) \in \hat{E} \implies (i,j) \in E \implies w_{ij}^- = 0 \leq 4b_{ij}$. Similarly, if $(i,j) \notin \hat{E}$ and $(i,j) \notin E$, then $w_{ij}^+ = b_{ij} = 0$. If $(i,j) \notin \hat{E}$ but $(i,j) \in E$, then $z_{ij} = b_{ij} \geq 1/2$ and so $w_{ij}^+ = 1 < 4b_{ij}$. 
	
	\noindent \textit{Checking condition 2.} For condition 2, note that $(i,j) \in \hat{E} \subseteq E$ and $(j,k) \in \hat{E}  \subseteq E$ imply that $z_{ij} + z_{jk} < 1$ and therefore $(i,j,k)$ is not an open wedge in $G$ and so $(i,k) \in E$. Since $(i,k) \notin \hat{E}$, we know $b_{ik} = z_{ik} \geq 1/2$. Overall, we have that
	\begin{equation*}
	w_{ij}^+ + w_{jk}^+ + w_{ik}^- = 2 = 4 \cdot \frac{1}{2} \leq 4b_{ik} \leq 4(b_{ij} + b_{jk} + b_{ik}).
	\end{equation*}
\end{proof}

\subsection{Proof of Theorem~\ref{thm:maince}}
\begin{proof}
	One convenient way to prove Theorem~\ref{thm:maince} is to show that it satisfies the conditions of Theorem~\ref{thm:3pt1} for an appropriate choice of weights, budgets, and parameter $\alpha$. 
	For cluster editing, recall that the correlation clustering weights are
	\begin{align*}
	(w_{ij}^+, w_{ij}^-) &=
	\begin{cases}
	(1,0) & \text{if $(i,j) \in E$} \\
	(0,1) & \text{if $(i,j) \notin E$}. \\
	\end{cases}
	\end{align*}
	For this theorem, we do not choose budgets to correspond to a lower bound for a labeling or clustering problems. Instead, the budgets are defined in terms of the set of flipped edges:
	\begin{align*}
	b_{ij} &= 
	\begin{cases}
	1 & \text{if $(i,j) \in E_W \cup E'$} \\
	0 & \text{otherwise}
	\end{cases}.
	\end{align*}
	The sum of budgets is exactly $\sum_{i< j} b_{ij} = |E'|+|E_W|$. Note then that the result holds if we can prove that the conditions of Theorem~\ref{thm:3pt1} are satisfied with $\alpha = 2$. 
	We first need to check that
	\begin{align}
	\label{first}
	&(i,j) \in \hat{E} \implies w_{ij}^- \leq 2b_{ij} \text{ and }\\
	\label{second}
	&(i,j) \notin \hat{E} \implies w_{ij}^+ \leq 2b_{ij}.
	\end{align}
	\textit{Checking~\eqref{first}:} If $(i,j) \in \hat{E}\cap E$ then $w_{ij}^- = 0 = b_{ij}$, and if $(i,j) \in \hat{E}$ but 
	$(i,j) \notin E$, then $b_{ij} = w_{ij}^- = 1$ since $(i,j)$ is a non-edge ($w_{ij}^- = 1$) that was flipped ($b_{ij} = 1$).
	
	\noindent\textit{Checking~\eqref{second}:} If $(i,j) \notin \hat{E}$ and $(i,j) \notin E$ then we have  $w_{ij}^+ = b_{ij} = 0$. If $(i,j) \notin \hat{E}$ and $(i,j) \in E$, then $w_{ij}^+ = 1 = b_{ij}$.

	Next we confirm that if $(i,j,k)$ is an open wedge centered at $j$ in $\hat{G} = (V, \hat{E})$, then
	\begin{align}
	w_{ij}^+ + w_{jk}^+ + w_{ik}^- \leq 2\left(   b_{ij} + b_{jk} + b_{ik} \right).
	\end{align}
	Regardless of the edge structure of $(i,j,k)$ in the original graph $G = (V,E)$, we must have
	\begin{equation}
	\label{sum3}
	b_{ij} + b_{jk} + b_{ik} + w_{ij}^+ + w_{jk}^+ + w_{ik}^- = 3.
	\end{equation}
	To see why, observe first of all that $(i,j) \in \hat{E}$, $(j,k) \in \hat{E}$, and $(i,k) \notin \hat{E}$, by our assumption that $(i,j,k)$ is an open wedge centered at $j$ in $\hat{G}$ . Consider node pair $(i,j)$: either this pair is an edge $(i,j) \in E$ (meaning $w_{ij}^+ = 1$) \emph{or} it was flipped (meaning $b_{ij} = 1$) but not both. Therefore, $b_{ij} + w_{ij}^+ = 1$, and by the same argument we can show $b_{jk} + w_{jk}^+ = b_{ik} + w_{ik}^- = 1$. This yields~\eqref{sum3}. 
	
	A key step in the proof is to realize that 
	\begin{equation}
	\label{sumb}
	b_{ij} + b_{jk} + b_{ik} \geq 1.
	\end{equation}
	If instead we assume $b_{ij} + b_{jk} + b_{ik} = 0$, this means that none of the edges were flipped, so $(i,j,k)$ is also an open wedge in the original graph $G = (V,E)$. This \emph{contradicts the fact that $(E',E_W)$ is a strong triadic closure labeling}. A strong triadic closure labeling would either add $(i,k)$ to the new edge set $E'$, or label one of the edges as weak, which would subsequently lead to one node pair being flipped. Combining~\eqref{sumb} and~\eqref{sum3}, we can see that 
	\begin{equation*}
	w_{ij}^+ + w_{jk}^+ + w_{ik}^- \leq 2 = 2(1) \leq 2 (b_{ij} + b_{jk} + b_{ik} ).
	\end{equation*}
\end{proof}

\subsection{Proof of Corollary~\ref{corce}}
\begin{proof}
	We have previously established that $\mathit{OPT}^+ \leq \mathit{OPT}^{CE}$. If $(E', E_W)$ is the $\alpha$-approximate \textsc{STC+} labeling returned by $\mathcal{A}$ and $B = |E'|+|E_W|$, then
	\begin{equation*}
	B \leq \alpha\mathit{OPT}^+ \leq \alpha\mathit{OPT}^{CE} \implies \frac{B}{\alpha} \leq  \mathit{OPT}^{CE},
	\end{equation*}
	which provides a lower bound on the optimal cluster editing solution. Using Theorem~\ref{thm:maince}, we can find a cluster editing solution that makes at most $2B$ mistakes, which is within $2\alpha$ of the lower bound. If we solve \textsc{minSTC+} optimally, this mean $\alpha = 1$, which shows that $\mathit{OPT}^+ \leq \mathit{OPT}^{CE} \leq 2\mathit{OPT}^+$.
\end{proof}
We also observe in passing that an $\alpha$-approximation algorithm for vertex cover would imply a $(2\alpha)$-approximation for cluster editing, since \textsc{minSTC+} can be reduced to vertex cover in an approximation preserving way. 

\subsection{Proof of Corollary~\ref{cor:6ce}}
\begin{proof}
	By construction, the minimum vertex cover in the 3-uniform hypergraph $\mathcal{H} = (V_\mathcal{H}, E_\mathcal{H})$ is equivalent to \textsc{minSTC+} on $G = (V,E)$. The algorithm performs the standard steps to obtain a 3-approximation: find a maximal matching, and place all nodes from the matched edges in the vertex cover. This can be converted to an \textsc{STC+} labeling that is a 3-approximation for \textsc{minSTC+}, which can be fed to \textsc{FlipPivot} to produce a $2\cdot3 = 6$ approximation for cluster editing.
\end{proof}

\subsection{Proof of Theorem~\ref{thm:maincd}}
\begin{proof}
	We must first confirm that this approach produces a feasible solution to cluster deletion, meaning that all clusters returned are cliques in the original graph $G = (V,E)$. Consider pivoting on any node $j$ in the derived graph $\hat{G} = (V,\hat{E})$. If $(j,k) \in \hat{E}$ and $(i,j) \in \hat{E}$, this means neither of these edges were labeled weak, and so we must have $(i,k) \in E$ or else strong triadic closure would be violated. Thus, pivoting on any node produces cliques.
	
	The rest of the theorem follows by showing that Theorem~\ref{thm:3pt1} holds with $\alpha = 2$ if we choose budgets $b_{ij} = 1$ if $(i,j) \in E_W$ and $b_{ij} = 0$ otherwise, and use weights $(w_{ij}^+, w_{ij}^-)$ that corresponding to cluster deletion (see~\eqref{weightscd}). In this case, the sum of budgets in is $\sum_{i<j} b_{ij} = |E_W|$.  
The conditions we must check are:
	\begin{enumerate}
		\item For all~$(i,j) \in \hat{E}$, we
		have~$w_{ij}^- \leq 2 b_{ij}$, and for all $(i,j) \notin \hat{E}$, we have $w_{ij}^+ \leq 2 b_{ij}$.
		\item If $(i,j,k)$ is an open wedge centered at $j$ in $\hat{G}$, we have $w_{ij}^+ + w_{jk}^+ + w_{ik}^- \leq 2 \left(   b_{ij} + b_{jk} + b_{ik} \right)$.
	\end{enumerate}
	\textit{Checking condition 1:} If $(i,j) \in \hat{E}$, then $(i,j) \in E$, so $w_{ij}^- = 0 \leq 2 b_{ij}$. If $(i,j) \notin \hat{E}$ and $(i,j) \notin E$, then have $w_{ij}^+ = 0 \leq 2b_{ij}$. If $(i,j) \notin \hat{E}$ but $(i,j) \in E$, then $(i,j) \in E_W$ and so $b_{ij} = 1$, and thus $w_{ij}^+ = 1 = b_{ij}$.
	
	\noindent \textit{Checking condition 2:} If $(i,j) \in \hat{E}$ and $(j,k) \in \hat{E}$, then we must have $(i,k) \in E$ or else there would be a violation of strong triadic closure. Since we are assuming in condition 2 that $(i,j,k)$ is an open wedge centered at $j$ in $\hat{G}$, the edge $(i,k) \in E_W$, and so $b_{ik} = 1$. Thus, we have
	\begin{equation*}
		w_{ij}^+ + w_{jk}^+ + w_{ik}^- = 2 = 2b_{ik}.
	\end{equation*}
	Therefore, the weight of mistakes (i.e, the number of deleted edges) resulting from running a pivoting procedure on $\hat{G} = (V, E-E_W)$ is at most $\alpha \sum_{i<j} b_{ij} = 2 |E_W|$ in expectation.
\end{proof}

We omit proofs for Corollaries~\ref{corcd} and~\ref{cor:4cd} as they follows the same arguments as Corollaries~\ref{corce} and~\ref{cor:6ce}.

\section{Deterministic Approximation Algorithms}
\label{app:det}
Our main text focused on randomized algorithms with expected approximation guarantees that can be obtained by applying a standard pivot procedure to a derived graph $\hat{G} = (V, \hat{E})$. All of our algorithms can be made deterministic by applying the deterministic pivoting procedure of van Zuylen and Williamson~\yrcite{vanzuylen2009deterministic}. Algorithm~\ref{alg:detpiv} provides pseudocode for this method, whose choice of pivot nodes is guided by the derived graph $\hat{G}$, the budgets $\{b_{ij}\}$, and the weights $\{w_{ij}^+, w_{ij}^-\}$ that define the correlation clustering instance. For completeness, we show how to set all parameters for our algorithms to obtain deterministic approximation guarantees.
\begin{algorithm}[tb]
	\caption{$\textsc{DetPivot}(V,W^+, W^-,\{b_{ij}\}, \hat{E})$}
	\begin{algorithmic}[5]
		\STATE{\bfseries Input:} Correlation clustering instance $(V, W^+, W^-)$, budgets $\{b_{ij}\}$, derived graph $\hat{G} = (V,\hat{E})$
		\STATE {\bfseries Output:} $\mathcal{C} =\textsc{DetPivot}(V,W^+, W^-,\{b_{ij}\}, \hat{E})$
		\FOR{$k \in V$}
		\STATE $T_k^+ =  \{ (i,j) \in \hat{E} \colon (j,k) \notin \hat{E}, (i,k) \in \hat{E} \}$ 
		\STATE  $T_k^- =  \{ (i,j) \notin \hat{E} \colon (j,k) \in \hat{E}, (i,k) \in \hat{E} \}$
		\STATE $P_k = 	{ \sum_{(i,j) \in T_k^+} w^+_{ij} + \sum_{(i,j)\in T_k^-} w_{ij}^-  \over \sum_{(i,j) \in T_k^+ \cup T_k^- } b_{ij} }$
		\ENDFOR
		\STATE $p = \argmin_{k \in V} P_k$  	\hfill	\texttt{// select pivot}
		\STATE $S = \{v \in V: (p,v) \in \hat{E} \}$ \hfill\texttt{// form cluster}
		\STATE $T = V \backslash S$ 	\hfill \texttt{// update node set}
		\STATE $W^+_T = \{w_{ij}^+ \colon i \in T, j \in T\}$ \hfill \texttt{// update remaining weights, edges, and budgets}
		\STATE $W^-_T = \{w_{ij}^- \colon i \in T, j \in T\}$
		\STATE $\hat{E}_T = \{(i,j) \in \hat{E} \colon i \in T, j\in T \}$
		\STATE $\mathcal{B}_T = \{b_{ij} \colon i \in T, j \in T \}$
		\STATE Return $\mathcal{C} = \{S, \textsc{DetPivot}(T,W^+_T, W^-_T,\mathcal{B}_T, \hat{E}_T)$
	\end{algorithmic}
	\label{alg:detpiv}
\end{algorithm}

Algorithms~\ref{alg:det4ce} and~\ref{alg:det4cd} are deterministic versions of Algorithms~\ref{alg:4owl} and~\ref{alg:4cd}, and show explicitly how to set budgets and weights to guide the deterministic pivoting strategy (Algorithm~\ref{alg:detpiv}).
\begin{algorithm}
	\caption{Deterministic Rounding for the \textsc{minSTC+} LP relaxation.}
	\begin{algorithmic}[5]
		\STATE{\bfseries Input:} Graph $G = (V,E)$
		\STATE {\bfseries Output:} Clustering of $G$.
		\STATE Solve LP-relaxation of~\eqref{owlx}
		\STATE Set $\hat{E} \gets \{(i,j) \in V \times V : x_{ij} < 1/2 \}$
		\STATE For $(i,j) \in E$, $(w_{ij}^+, w_{ij}^-) = (1,0)$, and $b_{ij} = x_{ij}$
		\STATE For $(i,j) \notin E$, $(w_{ij}^+, w_{ij}^-) = (0,1)$, and $b_{ij} = 1-x_{ij}$
		\STATE Return $\textsc{DetPivot}(V, \{w_{ij}^+\}, \{w_{ij}^-\},\{b_{ij}\}, \hat{E})$
	\end{algorithmic}
	\label{alg:det4ce}
\end{algorithm}
\begin{algorithm}
	\caption{Deterministic Rounding for the \textsc{minSTC} LP relaxation.}
	\begin{algorithmic}[5]
		\STATE{\bfseries Input:} Graph $G = (V,E)$
		\STATE {\bfseries Output:} Feasible cluster deletion clustering of $G$.
		\STATE Solve LP relaxation of~\eqref{minstc}
		\STATE Set $\hat{E} \gets \{(i,j) \in E : z_{ij} < 1/2 \}$
		\STATE For $(i,j) \in E$, $(w_{ij}^+, w_{ij}^-) = (1,0)$, and $b_{ij} = z_{ij}$
		\STATE For $(i,j) \notin E$, $(w_{ij}^+, w_{ij}^-) = (0,\infty)$ and $b_{ij} = 0$
		\STATE Return $\textsc{DetPivot}(V, \{w_{ij}^+\}, \{w_{ij}^-\},\{b_{ij}\}, \hat{E})$
	\end{algorithmic}
	\label{alg:det4cd}
\end{algorithm}
Similarly, Algorithm~\ref{alg:detmfpce} is a deterministic counterpart to Algorithm~\ref{alg:6ce}, and provides a way to turn an STC+ labeling $(E', E_W)$ into a cluster editing solution that is guaranteed to make at most $2(|E'|+ |E_W|)$ mistakes (i.e., deleted and added edges). Finally, Algorithm~\ref{alg:detmfpcd} is a deterministic version of Algorithm~\ref{alg:mfpcd}, our combinatorial 4-approximation for cluster deletion.
\begin{algorithm}
	\caption{$\textsc{DetMatchFlipPivotCE}(G,E',E_W)$}
	\begin{algorithmic}[5]
		\STATE{\bfseries Input:}  $G = (V,E)$ 
		\STATE {\bfseries Output:} Clustering of $G$.
		\STATE \textit{Reduce:} Build open wedge hypergraph $\mathcal{H} = (V_\mathcal{H}, E_\mathcal{H})$ (Section~\ref{sec:stcreductions})
		\STATE \textit{Match:} Find maximal matching $\mathcal{M} \subseteq E_\mathcal{H}$
		\STATE \textit{Vertex Cover:} ${C} = \{ v_{ij} \in V_\mathcal{H} \colon v_{ij} \in w \text{ for some  $w \in \mathcal{M}$} \}$
		\STATE \textit{STC+ Labeling}:
		\STATE\begin{center}
			$E' = \{ (i,j) \notin E \colon v_{ij} \in  {C} \}$\\
			$E_W = \{ (i,j) \in E \colon v_{ij} \in {C} \}$
		\end{center}
		\STATE Construct $\hat{G} = (V, \hat{E})$ where $\hat{E} = E' \cup (E - E_W)$
		\STATE Set budgets $(b_{ij})$ and weights $(w_{ij}^+, w_{ij}^-)$:
		\begin{align*}
		b_{ij} = 
		\begin{cases}
		1 & \text{if $(i,j) \in E_W \cup E'$} \\
		0 & \text{otherwise}
		\end{cases}
		\hspace{2cm}
		(w_{ij}^+, w_{ij}^-) =
		\begin{cases}
		(1,0) & \text{if $(i,j) \in E$} \\
		(0,1) & \text{if $(i,j) \notin E$} \\
		\end{cases}
		\end{align*}
		\STATE Return $\textsc{DetPivot}(V, \{w_{ij}^+\}, \{w_{ij}^-\},\{b_{ij}\}, \hat{E})$
	\end{algorithmic}
	\label{alg:detmfpce}
\end{algorithm}
\begin{algorithm}
	\caption{$\textsc{DetMatchFlipPivotCD}(G,E_W)$}
	\begin{algorithmic}[5]
		\STATE{\bfseries Input:}  Graph $G = (V,E)$ and \textsc{STC} label set $E_W$
		\STATE {\bfseries Output:} Feasible cluster deletion clustering of $G$.
		\STATE \textit{Reduce:} Build Gallai graph $\mathcal{G} = (V_\mathcal{G}, E_\mathcal{G})$ (Section~\ref{sec:stcreductions})
		\STATE \textit{Match:} Find maximal matching $\mathcal{M} \subseteq E_\mathcal{G}$
		\STATE \textit{Cover:} $\mathcal{C} = \{ v_{ij} \in V_\mathcal{G} \colon v_{ij} \in w \text{ for some $w \in \mathcal{M}$} \}$ 
		\STATE \textit{STC Labeling}: $E_W = \{ (i,j) \in E \colon v_{ij} \in  \mathcal{C} \}$
		\STATE Construct graph $\hat{G} = (V, \hat{E})$ where $\hat{E} = (E - E_W)$
		\STATE Set budgets $(b_{ij})$ and weights $(w_{ij}^+, w_{ij}^-)$:
		\begin{align*}
		b_{ij} = 
		\begin{cases}
		1 & \text{if $(i,j) \in E_W$} \\
		0 & \text{otherwise}
		\hspace{2cm}
		\end{cases} &(w_{ij}^+, w_{ij}^-) =
		\begin{cases}
		(1,0) & \text{if $(i,j) \in E$} \\
		(0,\infty) & \text{if $(i,j) \notin E$} \\
		\end{cases}
		\end{align*}
		\STATE Return $\textsc{DetPivot}(V, \{w_{ij}^+\}, \{w_{ij}^-\},\{b_{ij}\}, \hat{E})$
	\end{algorithmic}
	\label{alg:detmfpcd}
\end{algorithm}

\section{Extended Experimental Results and Details}

\subsection{Graph details}

Aside from PACE challenge graphs, all of the graphs we consider in our experimental results come from either the Facebook100 dataset~\cite{traud2012facebook}, or are available on the Suitesparse Matrix Collection~\cite{suitesparse2011davis}, or the SNAP large network repository~\cite{snapnets}. These can be categorized into the following classes:
\begin{itemize}
	\item \textbf{fb-social}: Facebook social networks (all graphs from the Facebook100 dataset)
	\item \textbf{loc-social}: Location-based social networks (loc-Gowalla, loc-Brightkite)
	\item \textbf{o-social}: Other social networks (com-LiveJournal, com-Youtube, soc-Epinions1, soc-LiveJournal1, soc-Slashdot0811, soc-Slashdot0902)	
	\item \textbf{web}: web networks (web-BerkStan, web-Google, web-NotreDame, web-Stanford, wiki-topcats)
	\item	\textbf{comm}: communication networks (email, email-Enron, email-EuAll, wiki-Talk)
	\item	\textbf{road}: road networks (roadNet-CA, roadNet-PA, roadNet-TX)
	\item	\textbf{prod}: product co-purchasing networks (amazon0302, amazon0312, amazon0505, amazon0601, com-Amazon)
	\item	\textbf{collab}: collaboration networks (Erdos991, Netscience, ca-AstroPh, ca-CondMat, ca-GrQc, ca-HepTh, ca-HepPh, condmat2005, com-DBLP)
	\item	\textbf{cit}: citation networks (SmaGri, cit-HepPh, cit-HepTh, cit-Patents)
	\item	\textbf{bio}: biological networks (celegans-neural, celegans-metabolic)
	\item \textbf{other}: all other graphs (Harvard500, Roget, polblogs)
\end{itemize}
Some of these graphs have weights or directions. Before running our experiments, we standardize all graphs by removing edge directions and weights. 

\subsection{Results for Cluster Deletion Approximation Algorithms}
In Table~\ref{tab:snapcd} we show more results for running our approximation algorithms for cluster deletion on various graphs. Overall, our LP-STC algorithm is twice as fast as the canonical relaxation algorithm (LP-CD), while obtaining similar approximation results. Our match-flip-pivot technique (MFP-CD) is far more scalable, and comes with a minor loss in approximation guarantees.

\subsection{Extended Results and Details for Cluster Editing Approximation Algorithms}
Solving cluster editing on a graph $G$ is equivalent to solving the complete unweighted correlation clustering objective on the signed graph obtained by treating edges of $G$ as positive edges and non-edges in $G$ as negative edges. Cluster editing is much more computationally expensive than cluster deletion, as it involves both deleting \emph{and adding} edges, rather than just deleting edges. 
Approximation algorithms and LP-rounding schemes for this problem are more abundant than approximation algorithms for cluster deletion. Therefore, for this problem we provide additional details on techniques for scaling LP algorithms as much as possible, and we also provide additional details on different methods for rounding the lower bounds we consider.  Interestingly, we find in practice that the best results are obtained by combining the lower bounds of our algorithms with alternative rounding schemes than the ones we theoretically analyze in the main text. This suggests that improved approximation results may also be obtained by analyzing other rounding strategies.

\paragraph{Rounding for MFP-CE}
For MFP-CE, we compare our combinatorial lower bound against three types of pivoting strategies. The first is the standard approach of applying a random \textsc{Pivot} procedure to the derived graph $\hat{G}$ in Algorithm~\ref{alg:6ce}. The second is a proof-of-concept implementation of the deterministic version of this algorithm (Algorithm~\ref{alg:detmfpce}), which significantly improves a posteriori approximation guarantees but is slower as our implementation of the deterministic pivoting procedure is not optimized. An optimized version of this code would be significantly faster, though we expect it to typically still be noticeably slower than randomized pivot node selections. Finally, we find that the best results can be obtained simply by running standard \textsc{Pivot} on the original graph $G$ multiple times and comparing the best result against the lower bound from MFP-CE. This last approach is extremely fast while producing results that are comparable to deterministic pivoting on the derived graph $\hat{G}$. In Tables~\ref{tab:snapcc_smaller} and~\ref{tab:snapcc_larger}, we show results for each rounding method (when applying randomized \textsc{Pivot}, we take the best result from 50 different runs). In Table~\ref{tab:ccshort} in the main text, we have displayed results for the standard MFP-CE algorithm that applies pivoting on the derived graph $\hat{G}$, in order to show how the approximation algorithm performs when it is run exactly according to its theoretical design. In subsequent experiments in the main text, we display results for applying \textsc{Pivot} to the original graph $G = (V,E)$, as this perform better typically and is just as fast.

\paragraph{Rounding for LP-STC+}
For the canonical LP-relaxation (LP-CE), we use the rounding scheme with a 2.06-approximation guarantee, due to Chawla, Makarychev, Schramm, and Yaroslavtsev~\yrcite{ChawlaMakarychevSchrammEtAl2015}, which we refer to as \textit{CMSY} rounding. This involves a more careful randomized construction of a derived graph before running a pivot procedure; we perform this randomized construction ten times and take the best result. For LP-STC+, we use the more simplistic rounding scheme that gives us our 4-approximation (Algorithm~\ref{alg:4owl}). For our results in the appendix, we additionally apply the same \textit{CMSY} rounding procedure, which is cheap in comparison with finding the lower bound. We find in practice that \textit{CMSY} produces the best results. 

\paragraph{Scalability and LP solvers.}
Linear programming relaxations for cluster deletion have $O(|E|)$ variables, whereas LPs for cluster editing involve $O(|V|^2)$ variables. Even more significantly, the canonical LP for cluster editing has $O(|V|^3)$ constraints. In our experiments, it becomes prohibitively expensive to even \emph{form} the constraint matrix when $|V|$ equals a few hundred. In the runtimes listed in Table~\ref{tab:ccshort} and in the appendix, the runtimes for LP-STC+ and LP-CE appear very similar, but this is only because we apply a useful warm-start approach for solving the canonical LP relaxation. This approach first solves the LP relaxation for STC+ and then iteratively adds in constraints from the canonical cluster editing relaxation that were violated. This process continues adding in violated constraints and re-solving the problem until all of the canonical LP constraints are satisfied, even if they were not included explicitly. Often, the solution for the STC+ relaxation matches the solution for LP-CE and this procedure terminates quickly, while in many other cases only a few iterations are needed. This lazy constraints approach has previously been used to help scale up LP solvers for correlation clustering~\cite{veldt2019metric}, though without an explicit realization that this method actually begins by solving the STC+ relaxation.

There also exist specialized solvers for \emph{approximately} solving the correlation clustering LP relaxation by applying memory-efficient projection methods~\cite{ruggles2020parallel,veldt2019metric,sonthalia2020project}. However, although these methods come with a smaller memory requirement and can be generalized to other weighted variants of the problem, they are still quite slow and were not competitive for the problems we considered. In Table~\ref{tab:metricopt} we show lower bounds and runtimes for using the projection method of~\citet{veldt2019metric} on a sample of graphs. When it comes to lower bounding the cluster editing relaxation, this method is much slower than \textsc{LP-STC+} and \textsc{LP-CE}, and returns poorer lower bounds.


\begin{table}[h!]
	\caption{
		Cluster editing lower bounds (LB) for three convex relaxations, and runtimes (Run). Value $n$ denotes number of nodes, $m$ is the number of edges. Proj-CE solves a quadratic program that approximates the canonical cluster editing LP. This method is not competitive with LP-STC+ or LP-CE when applied to cluster editing. Dashed lines indicate that Proj-CE did not converge after half an hour.
	}
	\label{tab:metricopt}
	\centering
	\scalebox{0.95}{\begin{tabular}{l l l l l l}
			\toprule
			\textbf{Graph} & && \textbf{LP-STC+} & \textbf{LP-CE} &\textbf{Proj-CE}   \\	
			\midrule
\textsc{Harvard500}&$n = 500$  & LB & 727.0 & 727.0 & 696.3 \\
& $m = 2043$ & Run & 0.376 & 0.447  & 33.4 \\
\midrule
\textsc{Roget}&$n = 994$  & LB & 1819.5 & 1819.5 & 1736.6 \\
& $m = 3640$ & Run & 0.622 & 0.893  & 218.8 \\
\midrule
\textsc{SmaGri}&$n = 1024$  & LB & 2457.0 & 2457.0 & 2345.3 \\
& $m = 4916$ & Run & 2.7 & 2.9  & 278.5 \\
\midrule
\textsc{polblogs}&$n = 1222$  & LB & 8356.0 & 8356.0 & 7976.3 \\
& $m = 16714$ & Run & 61.6 & 63.1  & 326.4 \\
\midrule
\textsc{email}&$n = 1133$  & LB & 2722.0 & 2722.0 & 2598.6 \\
& $m = 5451$ & Run & 1.6 & 1.9  & 416.0 \\
\midrule
\textsc{ca-GrQc}&$n = 5242$  & LB & 4931.0 & 4931.0 & -- \\
& $m = 14484$ & Run & 16.2 & 80.9  & --\\
\midrule
\textsc{caHepTh}&$n = 8638$  & LB & 11289.8 & 11290.5 & -- \\
& $m = 24806$ & Run & 64.7 & 625.5  & --\\
			\bottomrule
	\end{tabular}}
\end{table} 

\begin{table}
	\caption{
	Detailed cluster deletion results ($n = |V|$, and $m = |E|$). An asterisk next to the result of LP-STC indicates this LP relaxation returns an optimal solution for the canonical cluster deletion LP relaxation. Dashed lines indicate the method ran out of memory.
	}
	\label{tab:snapcd}
	\begin{minipage}[t]{0.49\textwidth}
		\centering
		\scalebox{0.95}{\begin{tabular}{lllll}
				\toprule
				\textbf{Graph} & & MFP-CD & LP-STC & LP-CD \\	
				\midrule
				\textsc{Netscience} & LB & 315 & 356.5$^*$ & 356.5 \\
				& UB & 669 & 597.0 & 605.0  \\
				$n = 379$ & Ratio & 2.124 & 1.675 & 1.697 \\
				$m = 914$ & Run & 0.00365 & 0.0117 & 0.0231 \\
				\midrule
				\textsc{Erdos991} & LB & 674 & 701.0$^*$ & 701.0 \\
				& UB & 1338 & 1372.0 & 1378.0  \\
				$n = 446$ & Ratio & 1.985 & 1.957 & 1.966 \\
				$m = 1413$ & Run & 0.000419 & 0.0513 & 0.101 \\
				\midrule
				\textsc{celegans-} & LB & 966 & 996.5 & 996.5  \\
				\textsc{metabolic} & UB & 1917 & 1961.0 & 1963.0  \\
				$n = 453$ & Ratio & 1.984 & 1.968 & 1.97 \\
				$m = 2025$ & Run & 0.00058 & 0.15 & 0.257 \\
				\midrule
				\textsc{Harvard500} & LB & 776 & 823.0$^*$ & 823.0 \\
				& UB & 1548 & 1584.0 & 1584.0  \\
				$n = 500$ & Ratio & 1.995 & 1.925 & 1.925 \\
				$m = 2043$ & Run & 0.00424 & 0.092 & 0.252 \\
				\midrule
				\textsc{celegans-} & LB & 1062 & 1074.0$^*$ & 1074.0 \\
				\textsc{neural}& UB & 2117 & 2148.0 & 2148.0  \\
				$n = 297$ & Ratio & 1.993 & 2.0 & 2.0 \\
				$m = 2148$ & Run & 0.000474 & 0.107 & 0.196 \\
				\midrule
				\textsc{Roget} & LB & 1788 & 1819.5$^*$ & 1819.5 \\
				& UB & 3571 & 3623.0 & 3624.0  \\
				$n = 994$ & Ratio & 1.997 & 1.991 & 1.992 \\
				$m = 3640$ & Run & 0.00117 & 0.141 & 0.143 \\
				\midrule
				\textsc{SmaGri} & LB & 2410 & 2457.0$^*$ & 2457.0 \\
				& UB & 4811 & 4909.0 & 4910.0  \\
				$n = 1024$ & Ratio & 1.996 & 1.998 & 1.998 \\
				$m = 4916$ & Run & 0.0015 & 0.435 & 0.58 \\
				\midrule
				\textsc{email} & LB & 2616 & 2722.0$^*$ & 2722.0 \\
				& UB & 5169 & 5430.0 & 5432.0  \\
				$n = 1133$ & Ratio & 1.976 & 1.995 & 1.996 \\
				$m = 5451$ & Run & 0.00248 & 0.234 & 0.478 \\
				\midrule
				\textsc{ca-GrQc} & LB & 4789 & 5196.0 & 5196.0  \\
				& UB & 10095 & 8598.0 & 8620.0  \\
				$n = 5242$ & Ratio & 2.108 & 1.655 & 1.659 \\
				$m = 14484$ & Run & 0.0674 & 0.559 & 2.2 \\
				\midrule
				\textsc{Caltech36} & LB & 8295 & 8324.5$^*$ & 8324.5 \\
				& UB & 16613 & 16648.0 & 16648.0  \\
				$n = 769$ & Ratio & 2.003 & 2.0 & 2.0 \\
				$m = 16656$ & Run & 0.101 & 4.36 & 9.56 \\
				\midrule
				\textsc{polblogs} & LB & 8336 & 8356.0$^*$ & 8356.0 \\
				& UB & 16660 & 16705.0 & 16706.0  \\
				$n = 1222$ & Ratio & 1.999 & 1.999 & 1.999 \\
				$m = 16714$ & Run & 0.00441 & 4.64 & 9.48 \\
				\midrule
				\textsc{Reed98} & LB & 9369 & 9405.5$^*$ & 9405.5 \\
				& UB & 18676 & 18811.0 & 18811.0  \\
				$n = 962$ & Ratio & 1.993 & 2.0 & 2.0 \\
				$m = 18812$ & Run & 0.00434 & 4.79 & 9.24 \\
				\bottomrule
		\end{tabular}}
		%
	\end{minipage}
	\begin{minipage}[t]{0.4\textwidth}
		\centering
		\scalebox{0.95}{\begin{tabular}{lllll}
				\toprule
				\textbf{Graph} & & MFP-CD & LP-STC & LP-CD  \\	
				\midrule
				\textsc{ca-HepTh} & LB & 10673 & 11320.5 & 11320.5  \\
				& UB & 21508 & 21631.0 & 21712.0  \\
				$n = 8638$ & Ratio & 2.015 & 1.911 & 1.918 \\
				$m = 24806$ & Run & 0.0848 & 1.42 & 2.38 \\
				\midrule
				\textsc{Simmons} & LB & 16451 & 16490.0$^*$ & 16490.0 \\
				& UB & 32919 & 32976.0 & 32976.0  \\
				$n = 1518$ & Ratio & 2.001 & 2.0 & 2.0 \\
				$m = 32988$ & Run & 0.02 & 11.5 & 18.5 \\
				\midrule
				\textsc{Haverford} & LB & 29771 & 29794.5$^*$ & 29794.5 \\
				& UB & 59508 & 59588.0 & 59588.0  \\
				$n = 1446$ & Ratio & 1.999 & 2.0 & 2.0 \\
				$m = 59589$ & Run & 0.0153 & 38.0 & 110.0 \\
				\midrule
				\textsc{Swarthmore} & LB & 30500 & 30524.0$^*$ & 30524.0 \\
				& UB & 61005 & 61048.0 & 61048.0  \\
				$n = 1659$ & Ratio & 2.0 & 2.0 & 2.0 \\
				$m = 61050$ & Run & 0.0292 & 36.0 & 68.4 \\
				\midrule
				\textsc{Bowdoin} & LB & 42163 & 42192.5$^*$ & 42192.5 \\
				& UB & 84298 & 84382.0 & 84382.0  \\
				$n = 2252$ & Ratio & 1.999 & 2.0 & 2.0 \\
				$m = 84387$ & Run & 0.0216 & 55.2 & 114.0 \\
				\midrule
				\textsc{Amherst} & LB & 45450 & 45477.0$^*$ & 45477.0 \\
				& UB & 90887 & 90953.0 & 90953.0  \\
				$n = 2235$ & Ratio & 2.0 & 2.0 & 2.0 \\
				$m = 90954$ & Run & 0.0227 & 70.0 & 162.0 \\
				\midrule
				\textsc{condmat05} & LB & 72428 & 79287.5 & 79287.5  \\
				& UB & 147826 & 152791.0 & 153446.0  \\
				$n = 36458$ & Ratio & 2.041 & 1.927 & 1.935 \\
				$m = 171734$ & Run & 0.433 & 39.5 & 72.5 \\
				\midrule
				\textsc{EmailEnron} & LB & 84385 & 87861.0 & 87861.0  \\
				& UB & 169793 & 173936.0 & 174035.0  \\
				$n = 36692$ & Ratio & 2.012 & 1.98 & 1.981 \\
				$m = 183831$ & Run & 0.398 & 243.0 & 391.0 \\
				\midrule
				\textsc{Rice31} & LB & 92342 & 92410.5$^*$ & -- \\
				& UB & 184692 & 184814.0 & --  \\
				$n = 4087$ & Ratio & 2.0 & 2.0 & -- \\
				$m = 184828$ & Run & 0.0952 & 227.0 & -- \\
				\midrule
				\textsc{ca-AstroPh} & LB & 87563 & 91188.0 & 91188.0  \\
				& UB & 178278 & 174918.0 & 174802.0  \\
				$n = 17903$ & Ratio & 2.036 & 1.918 & 1.917 \\
				$m = 196972$ & Run & 0.367 & 78.4 & 376.0 \\
				\midrule
				\textsc{Lehigh96} & LB & 99082 & 99172.5$^*$ & 99172.5 \\
				& UB & 198201 & 198345.0 & 198345.0  \\
				$n = 5075$ & Ratio & 2.0 & 2.0 & 2.0 \\
				$m = 198347$ & Run & 0.102 & 210.0 & 1420.0 \\
				\midrule
				\textsc{loc-}& LB & 101924 & 106429.0 & 106429.0  \\
				\textsc{Brightkite}& UB & 204104 & 211219.0 & 211240.0  \\
				$n = 58228$ & Ratio & 2.003 & 1.985 & 1.985 \\
				$m = 214078$ & Run & 0.632 & 151.0 & 241.0 \\
				\bottomrule
		\end{tabular}}
	\end{minipage}
\end{table}

\begin{table}
	\caption{
		Cluster editing (complete unweighted correlation clustering) results for smaller graphs. An asterisk next to the result of LP-STC+ indicates this LP relaxation returns an optimal solution for the canonical cluster editing LP relaxation for the given graph. Value $n$ denotes number of nodes, $m$ is the number of edges, LB is the lower bound returned by the method, UB is the upper bound achieved by rounding the lower bound, and Ratio = LB/UB is the a posteriori approximation guarantee. Runtime is given is seconds. We apply three ways to round the lower bound of MFP-CE. The first two are our own randomized ($+\mathit{piv}(\hat{G})$) and deterministic ($\mathit{det.}$, Algorithm~\ref{alg:detmfpce}) rounding strategies, which apply randomized and deterministic pivoting methods to a derived graph $\hat{G}$. The third ($+\mathit{piv}(G)$) is simply running \textsc{Pivot} on the original graph. For LP-STC+, we apply both our own randomized pivoting procedure ($+\mathit{piv}(\hat{G})$), to a derived graph $\hat{G}$ that differs from the derived graph generated by MFP-CE, as well as the rounding procedure that is guaranteed to output a 2.06 approximation (\textit{CMSY}) if applied to the cluster editing LP~\cite{ChawlaMakarychevSchrammEtAl2015}. The deterministic rounding for MFP-CE is slow as it is a proof-of-concept implementation, and not optimized.
	}
	\label{tab:snapcc_smaller}
	\centering
	\scalebox{0.95}{\begin{tabular}{ll | lll | ll |l}
			\toprule
			\textbf{Graph} & &\textbf{MFP-CE}  & & & \textbf{LP-STC+} &  & \textbf{LP-CE}  \\	
		 && $+\mathit{piv}(\hat{G})$ & $+\mathit{det.}$ & $+\mathit{piv}(G)$ & $+\mathit{piv}(\hat{G})$ & $+\mathit{CMSY}$ & $+\mathit{CMSY}$ \\
\midrule
\textsc{Harvard500} & LB & 687 & & & 727.0$^*$ & &727.0\\
& UB & 1832 & 1547 & 1331 & 1395 & 1303 & 1321  \\
$n = 500 $ & Ratio & 2.667 & 2.252 & 1.937 & 1.919 & 1.792 & 1.817  \\
$m = 2043$ & Run & 0.00333 & 0.0368 & 0.00381  & 0.419 & 0.426 & 0.451  \\
\midrule
\textsc{celegansneural} & LB & 1055 & & & 1074.0$^*$ & &1074.0\\
& UB & 2593 & 2341 & 2118 & 2148 & 2092 & 2100  \\
$n = 297 $ & Ratio & 2.458 & 2.219 & 2.008 & 2.0 & 1.948 & 1.955  \\
$m = 2148$ & Run & 0.012 & 0.0299 & 0.00435  & 0.58 & 0.583 & 0.59  \\
\midrule
\textsc{Roget} & LB & 1786 & & & 1819.5$^*$ & &1819.5\\
& UB & 4654 & 4061 & 3873 & 3621 & 3442 & 3413  \\
$n = 994 $ & Ratio & 2.606 & 2.274 & 2.169 & 1.99 & 1.892 & 1.876  \\
$m = 3640$ & Run & 0.00663 & 0.174 & 0.00949  & 0.632 & 0.656 & 0.844  \\
\midrule
\textsc{SmaGri} & LB & 2403 & & & 2457.0$^*$ & &2457.0\\
& UB & 6084 & 5330 & 5166 & 4905 & 4851 & 4817  \\
$n = 1024 $ & Ratio & 2.532 & 2.218 & 2.15 & 1.996 & 1.974 & 1.961  \\
$m = 4916$ & Run & 0.00845 & 0.282 & 0.0103  & 2.6 & 2.6 & 2.8  \\
\midrule
\textsc{email} & LB & 2606 & & & 2722.0$^*$ & &2722.0\\
& UB & 6729 & 5692 & 5859 & 5429 & 5269 & 5357  \\
$n = 1133 $ & Ratio & 2.582 & 2.184 & 2.248 & 1.994 & 1.936 & 1.968  \\
$m = 5451$ & Run & 0.00997 & 0.302 & 0.0105  & 1.4 & 1.4 & 1.7  \\
\midrule
\textsc{ca-GrQc} & LB & 4551 & & & 4931.0  & &4931.0  \\
& UB & 12414 & 9268 & 8610 & 8311 & 7595 & 7741  \\
$n = 5242 $ & Ratio & 2.728 & 2.036 & 1.892 & 1.685 & 1.54 & 1.57  \\
$m = 14484$ & Run & 0.0239 & 7.5 & 0.0358  & 15.4 & 15.9 & 73.4  \\
\midrule
\textsc{Caltech36} & LB & 8239 & & & 8324.5$^*$ & &8324.5\\
& UB & 18835 & 17204 & 17166 & 16648 & 17085 & 16891  \\
$n = 769 $ & Ratio & 2.286 & 2.088 & 2.084 & 2.0 & 2.052 & 2.029  \\
$m = 16656$ & Run & 0.0277 & 2.2 & 0.0167  & 118.9 & 118.9 & 119.1  \\
\midrule
\textsc{polblogs} & LB & 8309 & & & 8356.0$^*$ & &8356.0\\
& UB & 18869 & 17312 & 17664 & 16710 & 17212 & 17132  \\
$n = 1222 $ & Ratio & 2.271 & 2.084 & 2.126 & 2.0 & 2.06 & 2.05  \\
$m = 16714$ & Run & 0.0299 & 4.5 & 0.0182  & 63.9 & 64.0 & 64.4  \\
\midrule
\textsc{Reed98} & LB & 9338 & & & 9405.5$^*$ & &9405.5\\
& UB & 21805 & 19511 & 20386 & 18811 & 19723 & 19385  \\
$n = 962 $ & Ratio & 2.335 & 2.089 & 2.183 & 2.0 & 2.097 & 2.061  \\
$m = 18812$ & Run & 0.0311 & 2.5 & 0.0186  & 67.3 & 67.4 & 67.7  \\
\midrule
	\textsc{caHepTh} & LB & 10609 & & & 11289.8  & &11290.5  \\
& UB & 29049 & 23442 & 22214 & 21625 & 20470 & 20597  \\
$n = 8638 $ & Ratio & 2.738 & 2.21 & 2.094 & 1.915 & 1.813 & 1.824  \\
$m = 24806$ & Run & 0.049 & 12.0 & 0.0582  & 62.6 & 64.1 & 562.1  \\
			\bottomrule
	\end{tabular}}
\end{table}

\begin{table}
	\caption{
		Cluster editing (complete unweighted correlation clustering) results for larger graphs. Dashed lines indicate that the method ran out of memory. See caption of Table~\ref{tab:snapcc_smaller} for more information about each column and row meaning.
	}
	\label{tab:snapcc_larger}
	\centering
	\scalebox{0.95}{\begin{tabular}{ll | lll | ll |l}
			\toprule
			\textbf{Graph} & &\textbf{MFP-CE}  & & & \textbf{LP-STC+} &  & \textbf{LP-CE}  \\	
			&& $+\mathit{piv}(\hat{G})$ & $+\mathit{det.}$ & $+\mathit{piv}(G)$ & $+\mathit{piv}(\hat{G})$ & $+\mathit{CMSY}$ & $+\mathit{CMSY}$ \\
		\midrule
		\textsc{Simmons81} & LB & 16402 & & & 16490.0$^*$ & &16490.0\\
		& UB & 38856 & 34424 & 37919 & 32977 & 34374 & 34391  \\
		$n = 1518 $ & Ratio & 2.369 & 2.099 & 2.312 & 2.0 & 2.085 & 2.086  \\
		$m = 32988$ & Run & 0.0645 & 6.8 & 0.0263  & 235.8 & 235.9 & 236.9  \\
		\midrule
		\textsc{Haverford76} & LB & 29676 & & & --  & &--  \\
		& UB & 68742 & 61494 & 67437 & -- & -- & --  \\
		$n = 1446 $ & Ratio & 2.316 & 2.072 & 2.272 & -- & -- & --  \\
		$m = 59589$ & Run & 0.133 & 14.0 & 0.0611  & -- & -- & --  \\
		\midrule
		\textsc{Swarthmore42} & LB & 30411 & & & --  & &--  \\
		& UB & 70247 & 63138 & 69110 & -- & -- & --  \\
		$n = 1659 $ & Ratio & 2.31 & 2.076 & 2.273 & -- & -- & --  \\
		$m = 61050$ & Run & 0.125 & 16.3 & 0.0475  & -- & -- & --  \\
		\midrule
		\textsc{Bowdoin47} & LB & 42077 & & & --  & &--  \\
		& UB & 99161 & 87449 & 98881 & -- & -- & --  \\
		$n = 2252 $ & Ratio & 2.357 & 2.078 & 2.35 & -- & -- & --  \\
		$m = 84387$ & Run & 0.17 & 36.6 & 0.0668  & -- & -- & --  \\
		\midrule
		\textsc{Amherst41} & LB & 45346 & & & --  & &--  \\
		& UB & 105353 & 93857 & 103404 & -- & -- & --  \\
		$n = 2235 $ & Ratio & 2.323 & 2.07 & 2.28 & -- & -- & --  \\
		$m = 90954$ & Run & 0.226 & 39.4 & 0.0967  & -- & -- & --  \\
		\midrule
		\textsc{condmat2005} & LB & 71658 & & & --  & &--  \\
		& UB & 200854 & 151017 & 161977 & -- & -- & --  \\
		$n = 36458 $ & Ratio & 2.803 & 2.107 & 2.26 & -- & -- & --  \\
		$m = 171734$ & Run & 0.36 & 354.2 & 0.308  & -- & -- & --  \\
		\midrule
		\textsc{EmailEnron} & LB & 83991 & & & --  & &--  \\
		& UB & 221420 & 182653 & 184296 & -- & -- & --  \\
		$n = 36692 $ & Ratio & 2.636 & 2.175 & 2.194 & -- & -- & --  \\
		$m = 183831$ & Run & 0.428 & 828.4 & 0.334  & -- & -- & --  \\
		\midrule
		\textsc{Rice31} & LB & 92199 & & & --  & &--  \\
		& UB & 219129 & 189818 & 218071 & -- & -- & --  \\
		$n = 4087 $ & Ratio & 2.377 & 2.059 & 2.365 & -- & -- & --  \\
		$m = 184828$ & Run & 0.447 & 170.9 & 0.212  & -- & -- & --  \\
		\midrule
		\textsc{ca-AstroPh} & LB & 86369 & & & --  & &--  \\
		& UB & 225943 & 178356 & 186697 & -- & -- & --  \\
		$n = 17903 $ & Ratio & 2.616 & 2.065 & 2.162 & -- & -- & --  \\
		$m = 196972$ & Run & 0.488 & 496.8 & 0.302  & -- & -- & --  \\
		\midrule
		\textsc{Lehigh96} & LB & 98929 & & & --  & &--  \\
		& UB & 237145 & 204838 & 241219 & -- & -- & --  \\
		$n = 5075 $ & Ratio & 2.397 & 2.071 & 2.438 & -- & -- & --  \\
		$m = 198347$ & Run & 0.489 & 203.3 & 0.226  & -- & -- & --  \\
		\midrule
		\textsc{loc-Brightkite} & LB & 101544 & & & --  & &--  \\
		& UB & 267453 & 221993 & 239849 & -- & -- & --  \\
		$n = 58228 $ & Ratio & 2.634 & 2.186 & 2.362 & -- & -- & --  \\
		$m = 214078$ & Run & 0.559 & 1287.9 & 0.493  & -- & -- & --  \\
			\bottomrule
	\end{tabular}}
\end{table} 

\subsection{Details for Louvain Experiments and Cluster Deletion Results}
We used an implementation of the \textsc{LambdaLouvain} method from the author's previous work~\cite{veldt2020parameterized} (available at~\url{https://github.com/nveldt/ParamCC/blob/master/src/Graph_Louvain.jl}). Running this method with a parameter $\lambda = 1/2$ greedily optimizes the cluster editing objective, while $\lambda$ close to 1 greedily optimizes cluster deletion~\cite{Veldt:2018:CCF:3178876.3186110}.

The performance and runtime of this methods depends on how long the greedy procedure is allowed to run when searching for improved clusterings. In more detail, the method selects a random ordering of nodes, places all nodes in singleton clusters to start, and then iteratively visits each node to move it to the adjacent cluster leading to the greatest improvement in the correlation clustering objective. In theory, this greedy moving can continue until no more improvement is possible, but usually the number of passes over the nodes is truncated. Once there is no more improvement from visiting nodes, or once the maximum number of passes over the nodes is reached, the algorithm enters a second phase where nodes in the same cluster are agglomerated into supernodes and the procedure is run again on a reduced graph. Often, the algorithm is run multiple times with different random node orderings, and the best result is returned.

\paragraph{Parameter settings and cluster deletion results}
In our experiments, we run \textsc{LambdaLouvain} with the fastest possible settings. For each graph we fix a single random ordering of the nodes and run the algorithm only for this ordering. When iteratively visiting nodes to perform greedy moves, we visit each node only once time, and then skipped the second phase where nodes are agglomerated into supernodes. Since we use the fastest settings for \textsc{LambdaLouvain}, we do the same for MFP-CE, and only perform one step of the pivot rounding procedure for this comparison. This provides the most straightforward comparison between the methods, as we are comparing the fastest settings for each method. More importantly, we run \textsc{LambdaLouvain} with the fastest settings as this provides most of the improvement in terms of solution quality (i.e., the a posteriori approximation guarantee), at a fraction of the runtime. Overall this provides the most favorable comparison for \textsc{LambdaLouvain}. Even so, this method can be quite a bit slower than MFP-CE. Our plots do not report results for running \textsc{LambdaLouvain} on the largest SNAP graph, (soc-Livejournal1), as it took too long even with the fastest parameter settings.

\begin{figure}
	\centering
	\subfigure[SNAP CD  Runtimes] 
	{\includegraphics[width=.245\linewidth]{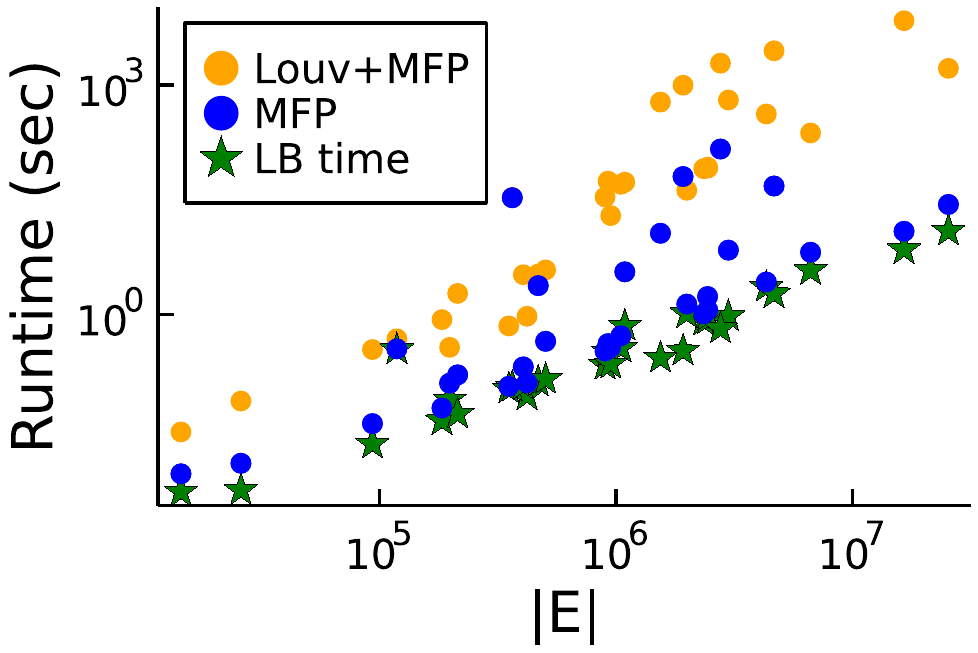}}\hfill
	\subfigure[FB100 CD Runtimes] 
	{\includegraphics[width=.245\linewidth]{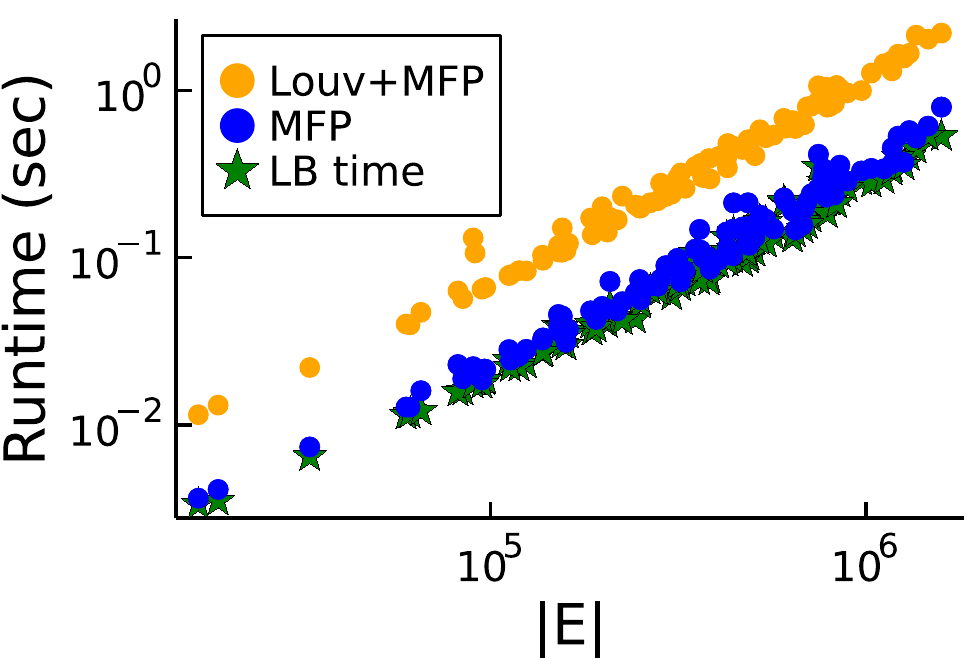}}\hfill
\subfigure[SNAP CD Ratios] 
{\includegraphics[width=.245\linewidth]{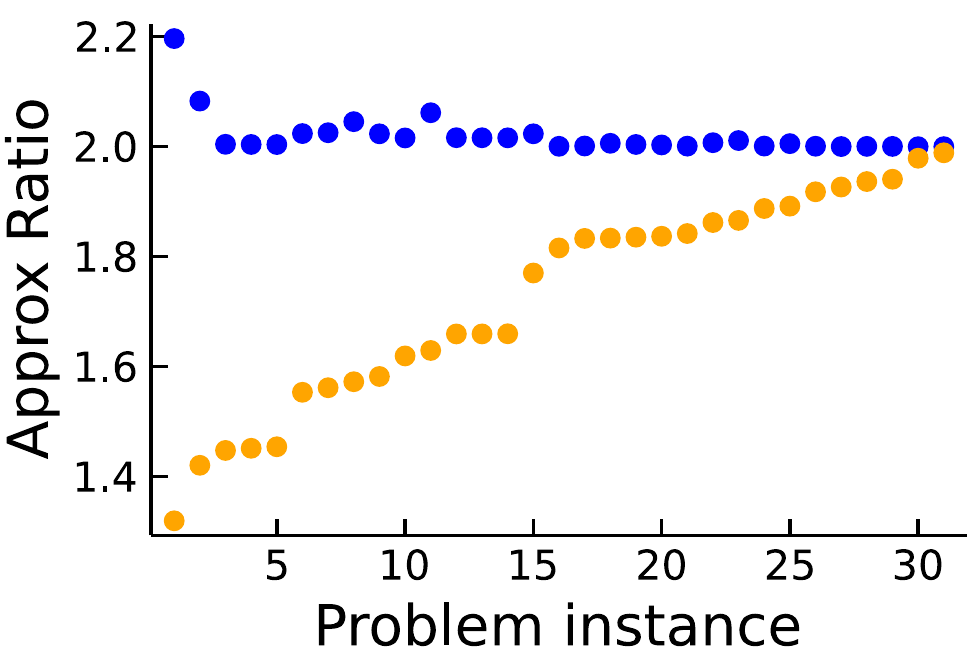}}\hfill
\subfigure[FB100 CD  Ratios]
{\includegraphics[width=.245\linewidth]{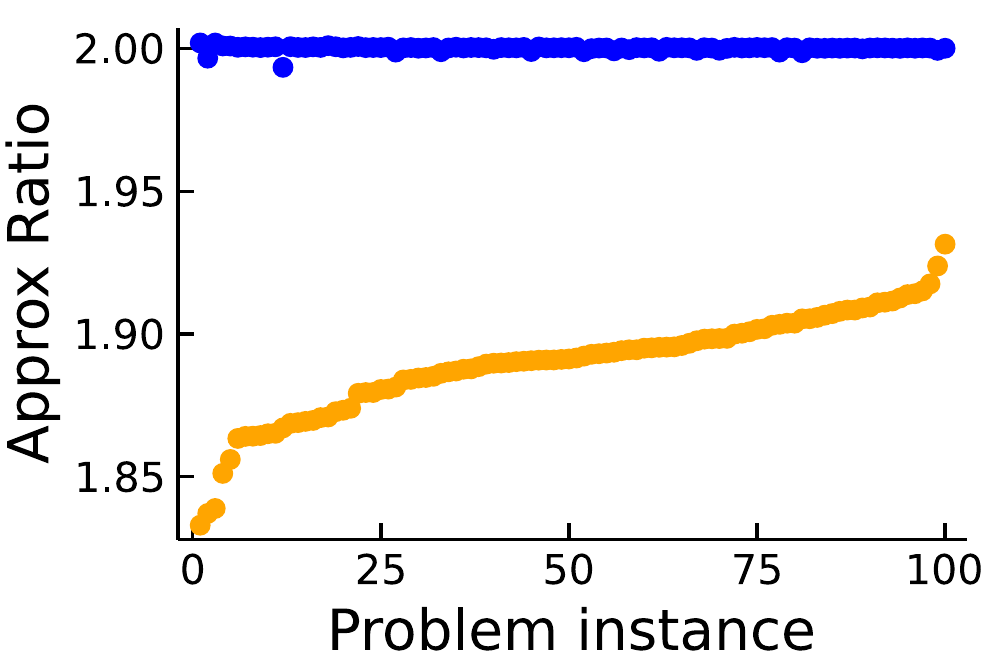}}\hfill
\vspace{-.5\baselineskip}
	\caption{Approximation ratios and runtimes when solving the cluster deletion objective.}
	\label{fig:louvaincd}
	\vspace{-\baselineskip} 
\end{figure}

We also provide approximation ratios and runtimes for MFP-CD and for \textsc{LambdaLouvain} when $\lambda$ is slightly less than one, which greedily optimizes the cluster deletion objective (Figure~\ref{fig:louvaincd}). MFP-CD tends to return solutions with an approximation guarantee around two for nearly every graph. \textsc{LambdaLouvain} again returns better solutions. This is often at the expense of longer runtimes, though for this objective MFP-CD is actually a little slower for some graphs. 
\begin{figure}
	\centering
	\subfigure[Alg Fastest (Runtime)] 
	{\includegraphics[width=.245\linewidth]{Figures/fb_runtime_ce_totaltime}}\hfill
	\subfigure[Alg Better Quality (Runtime)] 
	{\includegraphics[width=.245\linewidth]{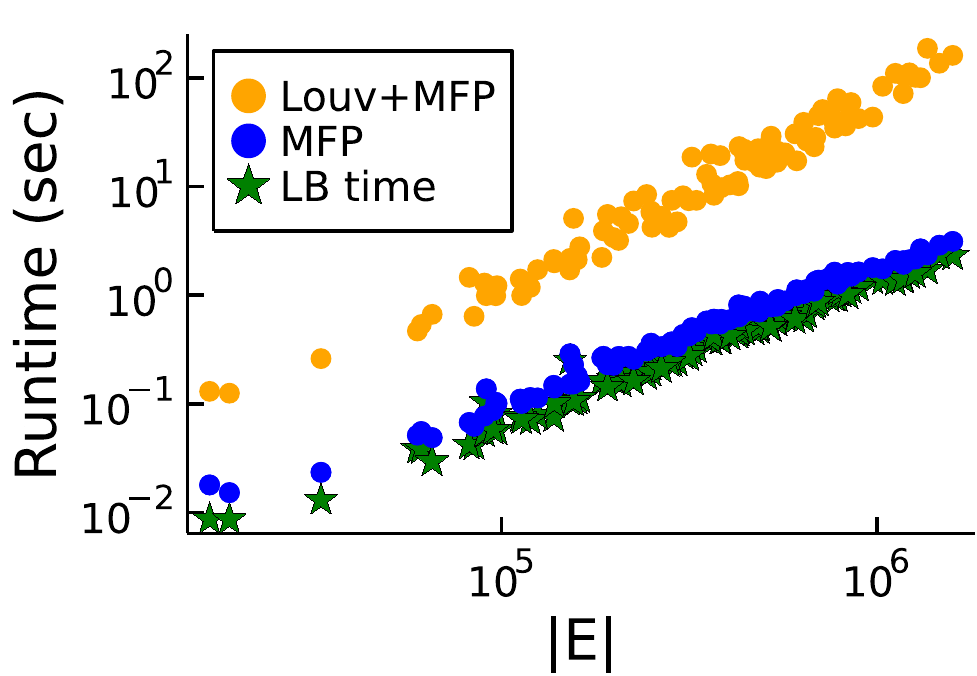}}\hfill
	\subfigure[Alg Fastest (Ratios)] 
	{\includegraphics[width=.245\linewidth]{Figures/fb_ratios_ce_totaltime}}\hfill
	\subfigure[Alg Better Quality (Ratios)] 
	{\includegraphics[width=.245\linewidth]{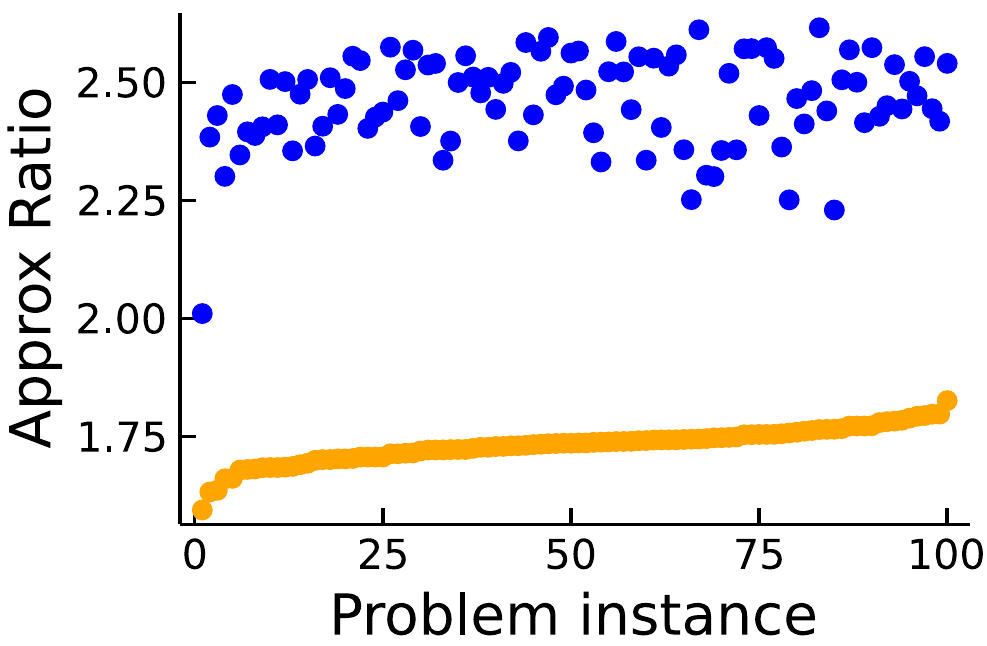}}\hfill
	\vspace{-.5\baselineskip}
	\caption{Approximation ratios and runtimes for MFP-CE and Louvain heuristics (Louv-MFP) on Facebook100 graphs. In (a) and (c), we use the fastest settings for each algorithm. For (b) and (d), we run the algorithms longer to obtain improved guarantees. Running the greedy Louvain heuristics longer leads to slightly improved guarantees at the expense of a significantly increased runtime.}
	\label{fig:fblonger}
	\vspace{-\baselineskip} 
\end{figure}

\paragraph{Better but slower results with Louvain} 
By visiting nodes more than once and performing the second phase of the Louvain method, we can obtain better results using \textsc{LambdaLouvain}, but this significantly increases the runtime. Meanwhile, for MFP-CE, increasing the number of pivot rounding steps has only a small increase in runtime since the main bottleneck for this method is the matching step for finding lower bounds. A more detailed comparison (specifically for the cluster editing objective) is provided in Figure~\ref{fig:fblonger}. Subfigures (a) and (c) are the same approximation and runtime plots from the main text, showing results for the fastest settings for each method. Subfigures (b) and (d) show results for running  \textsc{LambdaLouvain} when the number of passes over the nodes is increased to 10, and where we also apply the second phase of the algorithm. These plots also show results for running MFP-CE with 50 pivot steps for rounding. Running these algorithms longer leads to slight improvements to approximation ratios, though the results are qualitatively very similar. The runtime for MFP-CE is affected only slightly, but this increases the runtime for \textsc{LambdaLouvain} by an order of magnitude. 

\subsection{Additional Details for PACE Challenge Graphs}
Details and results from the original PACE challenge are available online at~\url{https://pacechallenge.org/2021/tracks/}. The KaPoCE algorithm was the winning method both for the exact and heuristic track of this challenge, so we focus on comparing against this method (\url{https://github.com/kittobi1992/cluster_editing}). The results displayed in the main text are for a subset of the benchmark graphs used for the heuristic track of the PACE challenge. In particular, we use odd-instance graphs from 1 to 57 (the even instance graphs were hidden and used as test cases for the challenge). These correspond to available benchmark graphs with fewer than 500 nodes. We capped the runtime of the heuristic version of KaPoCE to 10 minutes. Even for these small graphs, there were several cases where the algorithm reached this maximum runtime. The heuristic KaPoCE  algorithm obtains high quality solutions in terms of the cluster editing objective, but comes with no approximation guarantees of its own and does not scale to large graphs. Overall, our experiments illustrate that PACE challenge algorithms are simply focused on an alternative task---finding high quality solutions for small problems, rather than obtaining approximate solutions at a large scale.

\end{document}